\documentclass[aps,showpacs,prb,twocolumn,superscriptaddress]{revtex4}
\usepackage{amsmath}
\usepackage{graphicx}
\usepackage{dcolumn}
\usepackage{color}
\newcommand{\be}{\begin{equation}}
\newcommand{\ee}{\end{equation}}
\newcommand{\bea}{\begin{eqnarray}}
\newcommand{\eea}{\end{eqnarray}}
\newcommand{\bse}{\begin{subequations}}
\newcommand{\ese}{\end{subequations}}
\newcommand{\cuas}{${\rm EuCu_2As_2}$}
\newcommand{\cusb}{${\rm EuCu_2Sb_2}$}
\newcommand{\cusbB}{${\rm EuCu_{1.82}Sb_2}$}

\begin{document}
\title{Antiferromagnetism in ${\rm\bf EuCu_2As_2}$ and ${\rm\bf EuCu_{1.82}Sb_2}$ Single Crystals}
\author{V. K. Anand}
\altaffiliation{vivekkranand@gmail.com}
\affiliation {Ames Laboratory and Department of Physics and Astronomy, Iowa State University, Ames, Iowa 50011}
\affiliation{\mbox{Helmholtz-Zentrum Berlin f\"ur Materialien und Energie, Hahn-Meitner Platz~1, D-14109 Berlin, Germany}}
\author{D. C. Johnston}
\altaffiliation{johnston@ameslab.gov}
\affiliation {Ames Laboratory and Department of Physics and Astronomy, Iowa State University, Ames, Iowa 50011}

\date{\today}

\begin{abstract}

Single crystals of ${\rm EuCu_2As_2}$ and ${\rm EuCu_2Sb_2}$ were grown from CuAs and CuSb self-flux, respectively. The crystallographic, magnetic, thermal and electronic transport properties of the single crystals were investigated by room-temperature x-ray diffraction (XRD), magnetic susceptibility $\chi$ versus temperature $T$, isothermal magnetization $M$ versus magnetic field $H$, specific heat $C_{\rm p}(T)$ and electrical resistivity $\rho(T)$ measurements.  ${\rm EuCu_2As_2}$ crystallizes in the body-centered tetragonal ${\rm ThCr_2Si_2}$-type structure (space group $I4/mmm$), whereas ${\rm EuCu_2Sb_2}$ crystallizes in the related primitive tetragonal ${\rm CaBe_2Ge_2}$-type structure (space group $P4/nmm$). The  energy-dispersive x-ray spectroscopy and XRD data for the \cusb\ crystals showed the presence of vacancies on the Cu sites, yielding the actual composition \cusbB.  The $\rho(T)$ and $C_{\rm p}(T)$ data reveal metallic character for both ${\rm EuCu_2As_2}$ and \cusbB.  Antiferromagnetic (AFM) ordering is indicated from the $\chi(T)$, $C_{\rm p}(T)$, and $\rho(T)$ data for both ${\rm EuCu_2As_2}$ ($T_{\rm N} = 17.5$~K) and \cusbB\ ($T_{\rm N} = 5.1$~K).  In \cusbB, the ordered-state $\chi(T)$ and $M(H)$ data suggest either a collinear A-type AFM ordering of Eu$^{+2}$ spins $S=7/2$ or a planar noncollinear AFM structure, with the ordered moments oriented in the tetragonal $ab$~plane in either case. This ordered-moment orientation for the A-type AFM is consistent with calculations with magnetic dipole interactions.  The anisotropic $\chi(T)$ and isothermal $M(H)$ data for ${\rm EuCu_2As_2}$, also containing Eu$^{+2}$ spins $S=7/2$, strongly deviate from the predictions of molecular field theory for collinear AFM ordering and the AFM structure appears to be both noncollinear and noncoplanar.

\end{abstract}


\maketitle

\section{\label{Intro} Introduction}

The observation of high-$T_c$ superconductivity in FeAs-based 122-type compounds $A$Fe$_2$As$_2$ ($A$ = Ca, Sr, Ba) after suppressing the spin-density wave (SDW) transition is intriguing and these pnictides have been a topic of continuing research.\cite{Rotter2008a, Chen2008, Sasmal2008, Wu2008, Sefat2008, Torikachvili2008, Ishida2009, Johnston2010, Canfield2010, Mandrus2010} Because of their simple ${\rm ThCr_2Si_2}$-type body-centered tetragonal crystal structure and the availability of large single crystals, these compounds present a platform to understand the mechanism for the high-$T_c$ superconductivity and other interesting properties of the iron-arsenide class of superconductors. The $A$Fe$_2$As$_2$ ($A$ = Ca, Sr, Ba) compounds exhibit itinerant antiferromagnetic (AFM) spin-density wave (SDW) transitions that are accompanied by a tetragonal to orthorhombic structural distortion, and superconductivity is realized by partial substitution at the $A$, Fe and/or As~sites or by application of external pressure upon suppressing the SDW and structural transitions.\cite{Rotter2008a, Chen2008, Sasmal2008, Wu2008, Sefat2008, Torikachvili2008, Johnston2010} For example, K-doping at the Ba-site in BaFe$_2$As$_2$, which exhibits structural and SDW transitions near 140~K,\cite{Rotter2008b, Huang2008} suppresses these transitions and results in superconductivity with $T_c$ up to 38~K for $x \approx 0.4$ in Ba$_{1-x}$K$_x$Fe$_2$As$_2$.\cite{Rotter2008a} Cobalt~substitutions at the Fe~site of BaFe$_2$As$_2$ result in superconductivity with a maximum $T_c\sim25$~K at the optimum doping concentration $x \approx 0.06$ in Ba(Fe$_{1-x}$Co$_x$)$_2$As$_2$.\cite{Sefat2008,Ni2008, Wang2009} Isovalent P-doping at the As~site in BaFe$_2$As$_2$ also leads to superconductivity in BaFe$_2$(As$_{1-x}$P$_x$)$_2$ with $T_c \approx 30$~K for $x \approx 0.3$. \cite{Jiang2009a, Nakai2010} Superconductivity in BaFe$_2$As$_2$ can also be induced by application of external pressure.\cite{Alireza2009}

If nonmagnetic Ba$^{+2}$ in BaFe$_2$As$_2$ is completely replaced by Eu$^{+2}$ which carries a local moment with spin $S = 7/2$, an interesting situation occurs in which both itinerant conduction carrier magnetic moments and localized moments are present. In EuFe$_2$As$_2$, the Eu$^{+2}$ moments order antiferromagnetically below 19~K with an A-type AFM structure and the itinerant current carriers undergo a SDW transition at 190~K with an associated structural distortion.\cite{Ren2008, Tegel2008, Jiang2009c, Xiao2009} In the A-type structure, the Eu ordered moments are ferromagnetically aligned in each $ab$-plane layer with the ordered moments aligned in the $ab$~plane, but where the moments in adjacent layers along the $c$~axis are antiferromagnetically aligned (see Fig.~\ref{fig:structure}(b) below for the proposed A-type AFM structure of the Eu ordered moments in \cusb\ which is the same as in EuFe$_2$As$_2$).  The A-type AFM structure is therefore somewhat unusual, because the ferromagnetic (FM) alignment within an $ab$~plane often arises from dominant intraplane FM interactions, and the AFM then arises from weaker AFM interplane interactions.  The Curie-Weiss law for the magnetic susceptibility $\chi$ is
\begin{equation}
\chi(T) = \frac{C}{T-\theta_{\rm p}},
\label{eq:C-W}
\end{equation}
where $C$ is the Curie constant and $\theta_{\rm p}$ is the Weiss temperature. In the present context we consider a Heisenberg spin lattice with Hamiltonian ${\cal H} = \sum_{<ij>}J_{ij}{\bf S}_i\cdot {\bf S}_j$, where the sum is over distinct \{${\bf S}_i,\ {\bf S}_j$\} spin pairs and the exchange interactions $J_{ij}$ are positive for AFM interactions.  For a spin lattice containing identical crystallographically-equivalent spins such as the Eu$^{+2}$ spins $S=7/2$ in EuFe$_2$As$_2$ (and in ${\rm EuCu_2As_2}$ and \cusbB, see the following), the Weiss temperature is given in general by molecular field theory (MFT) as\cite{Johnston2011, Johnston2012, Johnston2015}
\be
\theta_{\rm p} = -\frac{S(S+1)}{3k_{\rm B}}\sum_jJ_{ij},
\ee
where $k_{\rm B}$ is Boltzmann's constant and the sum is over all neighboring spins~$j$ of a given central spin~$i$ with exchange interactions $J_{ij}$, respectively.  Thus if the dominant interactions are FM (negative), then $\theta_{\rm p}$ is positive (ferromagneticlike) as in EuFe$_2$As$_2$,\cite{Jiang2009c} even though EuFe$_2$As$_2$ is an antiferromagnet.  In this regard, we note that the similar compound EuCu$_2$P$_2$ does order ferromagnetically.\cite{Huo2011}  The Eu spins-7/2 in ${\rm EuCo_2As_2}$ were claimed to exhibit A-type AFM ordering below 39~K.\cite{Ballinger2012}

Like BaFe$_2$As$_2$, superconductivity is observed in EuFe$_2$As$_2$ after suppression of the SDW and the associated structural transition. The interaction of the Eu moments with superconductivity in pure and doped EuFe$_2$As$_2$ has been extensively studied. \cite{Jeevan2008, Anupam2009, Ren2009a, Miclea2009, Jiang2009b, Jeevan2011, Anupam2011} Partial substitution of Eu by K leads to superconductivity in Eu$_{1-x}$K$_{x}$Fe$_2$As$_2$ with $T_c$ as high as 33~K for $x =0.5$\@. \cite{Jeevan2008, Anupam2009, Anupam2011}  Re-entrant superconductivity is observed in EuFe$_2$As$_2$ on application of hydrostatic pressure. \cite{Miclea2009} Like substitution of Co for Fe in BaFe$_{2}$As$_{2}$, Co substitution for Fe in EuFe$_{2}$As$_{2}$ also leads to superconductivity but the superconductivity in Eu(Fe$_{1-x}$Co$_{x}$)$_{2}$As$_{2}$ is re-entrant, revealing an important role of the Eu magnetic moment.\cite{Jiang2009b} However, Ni substitution for Fe in EuFe$_{2}$As$_{2}$ does not induce superconductivity down to 2~K (Ref.~\onlinecite{Ren2009b}) which contrasts with the observation of superconductivity in Ni-doped BaFe$_{2}$As$_{2}$.\cite{Li2009,Ni2010} Interestingly, Ni-doped EuFe$_{2}$As$_{2}$ is reported to exhibit FM ordering of the Eu moments below 20 K,\cite{Ren2009b} which is not too surprising in view of the dominant FM interactions in EuFe$_2$As$_2$ as discussed above.  It was observed that partial subsitution of Ni for Fe in Eu$_{0.5}$K$_{0.5}$Fe$_2$As$_2$ leads to a revival of the SDW transition in ${\rm Eu_{0.5}K_{0.5}(Fe}_{1-x}{\rm Ni}_x)_2{\rm As}_2$ as the holes in the system created by K-doping at the Eu~site are compensated by the doped electrons introduced by partially replacing Ni for Fe.\cite{Anupam2012}

Recently we reported the crystallographic and physical properties of ${\rm SrCu_2As_2}$ and ${\rm SrCu_2Sb_2}$ which are $sp$-band metals. \cite{Anand2012} The Cu atoms are in a nonmagnetic 3$d^{10}$ electronic configuration with a formal Cu$^{+1}$ oxidation state. Here we report the physical properties of magnetic analogues of these compounds, namely ${\rm EuCu_2As_2}$ and ${\rm EuCu_2Sb_2}$. ${\rm EuCu_2As_2}$ crystallizes in the body-centered tetragonal (bct) ${\rm ThCr_2Si_2}$-type structure (space group $I4/mmm$) and ${\rm EuCu_2Sb_2}$ in the primitive tetragonal ${\rm CaBe_2Ge_2}$-type ($P4/nmm$) structure.\cite{Dunner1995}  D\"{u}nner~et~al.\cite{Dunner1995} reported the occurrence of vacancies on the Cu sites in ${\rm EuCu_2As_2}$ as deduced from their refinement of x-ray diffraction (XRD) data, yielding an actual composition ${\rm EuCu_{1.760(5)}As_2}$. No such vacancies were reported for ${\rm EuCu_2Sb_2}$.  The Eu atoms are crystallographically equivalent in both compounds.  Sengupta et al.\cite{Sengupta2005, Sengupta2012} reported magnetic and other physical properties of polycrystalline ${\rm EuCu_2As_2}$.  They found from $\chi(T)$ measurements that polycrystalline ${\rm EuCu_2As_2}$ orders antiferromagnetically below $T_{\rm N} = 15$~K, and from $\chi(T)$ and $^{151}$Eu M\"ossbauer measurements they found that the Eu ions are divalent ($S=7/2$).\cite{Sengupta2005}  From the temperature of the peak of the zero-field-cooled $\chi$ versus $T$ at fixed $H$ they inferred an approximately linear decrease in $T_{\rm N}$ with increasing $H$, where $T_{\rm N}(H=0)=15.5$~K and $T_{\rm N}\to0$ at $H=1.81$~T.\cite{Sengupta2012}  From electrical resistivity $\rho(T)$ measurments under pressure, they also inferred that $T_{\rm N}$ strongly increases with increasing pressure, reaching 49~K at a pressure of 10.7~GPa.\cite{Sengupta2012} The crystallographic and magnetic properties of the related compounds ${\rm EuPd_2As_2}$ (Ref.~\onlinecite{Anand2014}) and ${\rm EuPd_2Sb_2}$ (Ref.~\onlinecite{Das2010}) have also been reported, where AFM ordering of the Eu spins~7/2 is also found in each compound.  The detailed physical properties of ${\rm EuCu_2Sb_2}$ have not been reported before to our knowledge.

Herein, we report the growth of ${\rm EuCu_2As_2}$ and \cusbB\ single crystals and their crystallographic, magnetic, thermal and electronic transport properties investigated using powder XRD, $\chi(T)$, magnetization $M$ versus applied magnetic field $H$ isotherms, specific heat $C_{\rm p}(T)$, and electrical resistivity $\rho (T)$ measurements.  The experimental details are given in Sec.~\ref{ExpDetails} and the crystallographic results in  Sec.~\ref{Crystallography}.  The physical property measurements of \cusbB\ and ${\rm EuCu_2As_2}$ are presented and analyzed in Secs.~\ref{EuCu2Sb2} and \ref{EuCu2As2}, respectively. A summary is given in Sec.~\ref{Conclusion}.

\section{\label{ExpDetails} Experimental Details}

Single crystals of ${\rm EuCu_2As_2}$ and nominal ${\rm EuCu_2Sb_2}$ were grown by the self-flux method. The high-purity elements Eu (Ames Lab), Cu 99.999\% (Alfa Aesar),  As 99.99999\% (Alfa Aesar) and Sb 99.999\% (Alfa Aesar) were used for synthesis. Eu and prereacted flux (CuSb or CuAs) taken in a 1:5 molar ratio were placed in alumina crucibles which were then sealed in evacuated quartz tubes. The single crystals were grown by heating the quartz ampoules to 850~$^\circ$C at a rate of 60~$^\circ$C/h, held for 5~h, heated to 1100~$^\circ$C at a rate of 60~$^\circ$C/h, held for 25~h and then slowly cooled to 800~$^\circ$C at a rate of 2.5~$^\circ$C/h.  The crystals were separated from the flux by centrifugation at that temperature. We obtained shiny plate-like crystals of typical size $2.5 \times 2 \times 0.3$~mm$^3$ for ${\rm EuCu_2As_2}$ and $4 \times 3 \times 0.5$~mm$^3$ for ${\rm EuCu_2Sb_2}$.

The phase purity and the crystal structure of the crystals were determined by powder XRD collected on crushed single crystals using Cu K$\alpha$ radiation on a Rigaku Geigerflex x-ray diffractometer. The single-phase nature of the crystals was further checked using a JEOL scanning electron microscope (SEM) equipped with an energy-dispersive x-ray spectroscopy (EDS) analyzer. The single-phase nature of the crystals was inferred from the high resolution SEM images. Wavelength-dispersive x-ray spectroscopy (WDS) measurements were also carried out to determine the compositions of the crystals.  The average compositions obtained from the EDS and WDS measurements showed the expected 1:2:2 stoichiometry for \cuas\ and \cusb\ except for $\approx 9$\% vacancies on the Cu sites in \cusb\ corresponding to an actual composition ${\rm EuCu_{1.83(5)}Sb_2}$.  This latter result is consistent with the site occupancies determined from Rietveld refinements of the XRD patterns in Sec.~\ref{Crystallography} where a composition ${\rm EuCu_{1.82(1)}As_2}$ is found.  This composition is contrary to the full occupancies reported in Ref.~\onlinecite{Dunner1995}.  The difference likely arises from differences in sample preparation conditions. All molar magnetic and heat capacity measurements reported here for \cusbB\ are normalized to a mole of \cusbB\ formula units.

The magnetization measurements were performed using a Quantum Design, Inc., superconducting quantum interference device magnetic properties measurement system (MPMS).  The output of the MPMS software for the magnetic moment of a sample is in Gaussian cgs electromagnetic units (emu), which in terms of fundamental quantites is given by 1~emu = 1 G~cm$^3$.  The magnetic field unit is the Oe, with 1 Oe = 1 G, and we use the Tesla (T) as a unit of convenience where 1~T $= 10^4$~Oe.

Due to the large Eu spins-7/2 and the resulting demagnetizing fields that can be especially large at low temperatures, the  magnetic field~$H_{0\alpha}$ applied in the $\alpha^{\rm th}$ direction has been corrected for the demagnetizing field $H_{{\rm d}\alpha}$, yielding the internal field $H_{\rm int}$ in cgs units obtained from
\be
H_{{\rm int}\,\alpha} = H_{0\alpha} -4\pi N_{{\rm d}\alpha}M_\alpha,
\label{Eq:Hin0}
\ee
where the demagnetizing factor $N_{{\rm d}\alpha}$ is defined as for the SI system of units for which $0\leq N_{{\rm d}\alpha} \leq 1$ and $\sum_{\alpha=1}^3N_{{\rm d}\alpha} = 1$.  Thus the $M(H)$ isotherms are presented as $M$ versus $H_{\rm int}$ isotherms and the reported intrinsic $\chi_\alpha = M_\alpha/H_{\rm int}$ data are obtained from the observed $\chi^{{\rm obs}_\alpha} = M^{{\rm obs}_\alpha}/H_{0\alpha}$ data using 
\be
\chi_\alpha = \frac{\chi^{\rm obs}_\alpha} {1-4\pi N_{{\rm d}\alpha}\chi^{\rm obs}_\alpha}.
\label{Eq:ChiCorr0}
\ee
Since our crystal shapes can be approximated by rectangular prisms, the magnetometric $N_{{\rm d}\alpha}$ values were obtained using Eq.~(1) in Ref.~\onlinecite{Aharoni1998}.  Our reported magnetizations and magnetic susceptiblities are normalized to a mole of formula units (f.u.), whereas $M_{\alpha}$ in Eq.~(\ref{Eq:Hin0}) and $\chi^{\rm obs}_{\alpha}$ in the denominator of Eq.~(\ref{Eq:ChiCorr0}) are normalized to unit volume with respective units of G and dimensionless, respectively.  In cgs units the volume-normalized quantities are obtained from the ones normalized per mole of f.u.\ by dividing by the molar volume $V_{\rm M}[{\rm cm^3/mol}]$ as given in Table~\ref{tab:XRD} below.

The heat capacity was measured by a relaxation technique using a Quantum Design, Inc., physical properties measurement system (PPMS). The electrical resistivity measurements were performed using the standard four-probe ac technique using the ac-transport option of the PPMS with the current in the $ab$~plane. Annealed platinum wire ($50~\mu$m diameter) electrical leads were attached to the crystals using silver epoxy.

\begin{figure}
\includegraphics[width=3in]{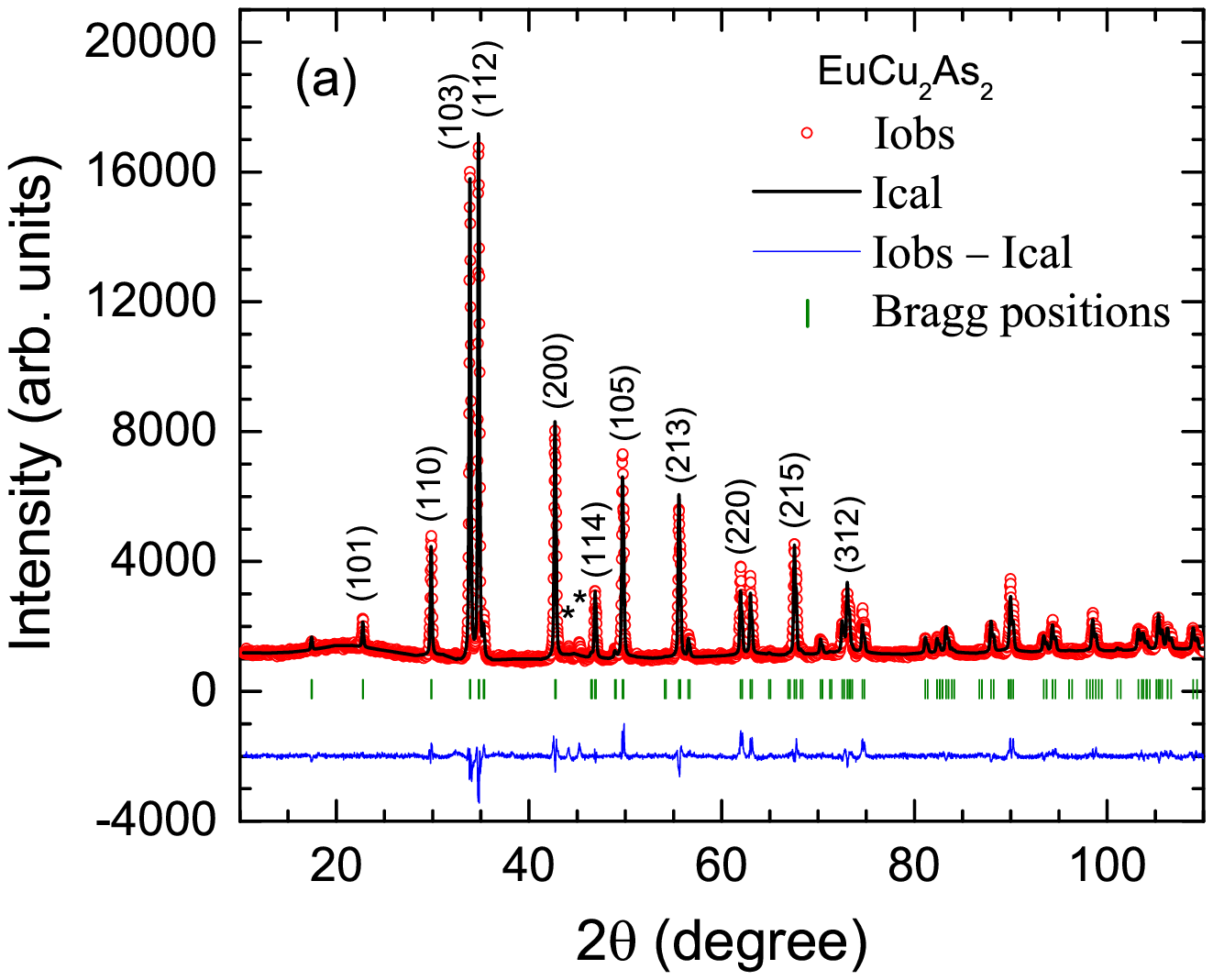}\vspace{0.1in}
\includegraphics[width=3in]{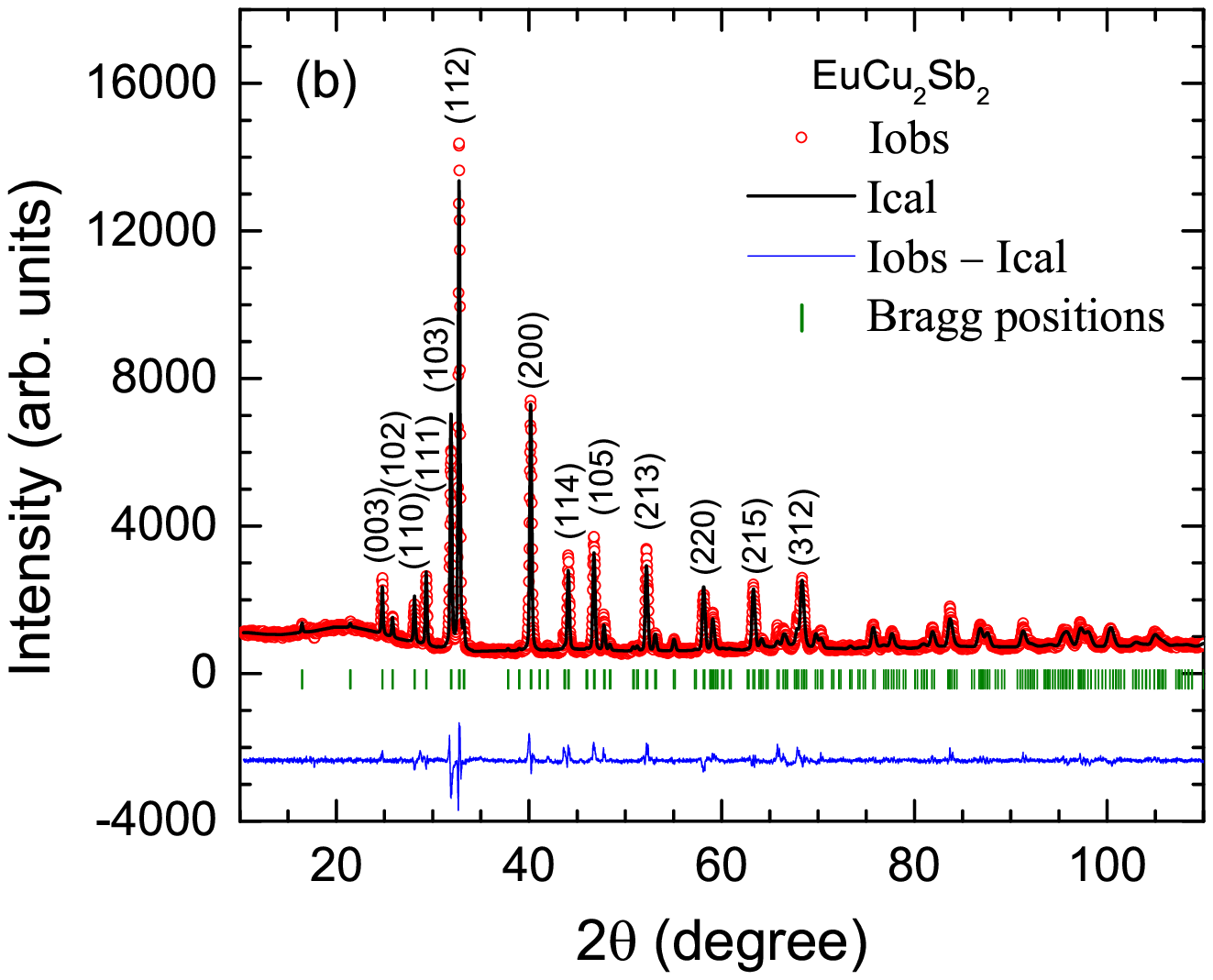}
\caption {(Colour online) Powder x-ray diffraction patterns of (a) ${\rm EuCu_2As_2}$ and (b) ${\rm EuCu_2Sb_2}$ recorded at room temperature. The solid line through the experimental points in (a) is the Rietveld refinement profile calculated for the ${\rm ThCr_2Si_2}$-type body-centered tetragonal structure (space group $I4/mmm$), and in (b) the Rietveld refinement profile for the  ${\rm CaBe_2Ge_2}$-type primitive tetragonal structure (space group $P4/nmm$). The short vertical bars mark the Bragg peak positions. The lowermost curves represent the differences between the experimental and calculated intensities. The unindexed peaks marked with stars correspond to peaks from the flux.  The Miller indices (hk$\ell$)of the strongest peaks in (a) and~(b) are indicated.  Whereas all (hk$\ell$) combinations are allowed for the primitive tetragonal structure of \cusb\ in (b), only indices with $h+k+l=$~even are allowed for the body-centered-tetragonal structure of \cuas\ in~(a).}
\label{fig:XRD}
\end{figure}

\begin{figure}
\includegraphics[width=3.3in]{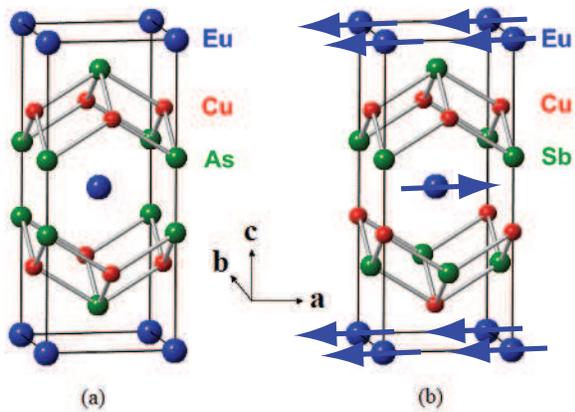}
\caption{(Color online) (a) ${\rm ThCr_2Si_2}$-type body-centered tetragonal crystal structure ($I4/mmm$) of ${\rm EuCu_2As_2}$. (b) ${\rm CaBe_2Ge_2}$-type primitive tetragonal crystal structure ($P4/nmm$) of ${\rm EuCu_2Sb_2}$. The order of the Cu and Sb layers in the lower half of the unit cell is reversed in (b) with respect to (a). In order to compare the structures, the origin of the EuCu$_2$Sb$_2$ unit cell is shifted by (1/4, 1/4, 1/4) from the atomic coordinates in Table~\ref{tab:XRD-coord}.  The arrows in (b) show the ordered Eu moments in an A-type AFM structure, which is one of two AFM structures proposed in this paper for \cusbB.  The other is a planar helix structure with the helix axis along the $c$~axis.}
\label{fig:structure}
\end{figure}

\section{\label{Crystallography} Crystallography}

The powder XRD data were collected on crushed single crystals of ${\rm EuCu_2Sb_2}$ and ${\rm EuCu_2As_2}$ at room temperature as shown in Fig.~\ref{fig:XRD}.  Also shown are the structural Rietveld refinement profiles of the XRD data using {\tt FullProf} \cite{Rodriguez1993} software. The refinements of the XRD data indicate that the crystals are single-phase since no extra peaks beyond those of the respective 122-type phases were observed except for the weak unindexed peaks marked with stars in the XRD pattern of ${\rm EuCu_2As_2}$ that arise from a small amount of adventitious flux on the surfaces of the crystals. Our Rietveld refinement confirmed the ${\rm ThCr_2Si_2}$-type bct structure ($I4/mmm$) for ${\rm EuCu_2As_2}$, and the ${\rm CaBe_2Ge_2}$-type primitive tetragonal crystal structure ($P4/nmm$) for ${\rm EuCu_2Sb_2}$. The two crystal structures are shown in Fig.~\ref{fig:structure}, and are both ternary derivatives of the binary ${\rm BaAl_4}$-type structure,\cite{Parthe1983} consisting of layers of Eu, Cu and As/Sb atoms stacked along the tetragonal $c$~axis.  From Fig.~\ref{fig:structure}, these two structures differ in the arrangement of layers of Cu and As/Sb layers. The order of the Cu and Sb layers in the lower half of the unit cell is reversed in the ${\rm CaBe_2Ge_2}$-type structure of ${\rm EuCu_2Sb_2}$ in Fig.~\ref{fig:structure}(b) compared to that in ${\rm ThCr_2Si_2}$-type ${\rm EuCu_2As_2}$ in Fig.~\ref{fig:structure}(a), thus resulting in a loss of the mirror plane perpendicular to the $c$~axis and of the inversion symmetry about the unique Eu position in the ${\rm CaBe_2Ge_2}$-type structure.  While all Cu atoms occupy equivalent Wyckoff $4d$ positions in the ${\rm ThCr_2Si_2}$ structure of ${\rm EuCu_2As_2}$, in the ${\rm CaBe_2Ge_2}$ structure of ${\rm EuCu_2Sb_2}$ the Cu atoms are equally distributed between the $2a$ and $2c$ positions. Thus in ${\rm EuCu_2Sb_2}$ there are two distinct types of Cu square lattices which have different lattice parameters and are rotated by 45$^\circ$ with respect to each other, in contrast to only one type of stacked square lattice of Cu atoms in ${\rm EuCu_2As_2}$.

\begin{table}
\caption{\label{tab:XRD} Crystallographic and Rietveld refinement parameters obtained from powder XRD data for crushed crystals of body-centered tetragonal ${\rm ThCr_2Si_2}$-type ${\rm EuCu_2As_2}$ (space group $I4/mmm$)  and primitive tetragonal ${\rm CaBe_2Ge_2}$-type \cusb\ (space group $P4/nmm$). The molar volume $V_{\rm M}$ (volume per mole of formula units) is also listed.  Shown for comparison are the lattice parameters for single crystals from Ref.~\onlinecite{Dunner1995} and for a polycrystalline sample of ${\rm EuCu_2As_2}$ from Ref.~\onlinecite{Sengupta2005}.}
\begin{ruledtabular}
\begin{tabular}{lccc}
& ${\rm EuCu_2As_2}$  \\
Lattice parameters						& \underline{This Work} 	&	\underline{Ref.~\onlinecite{Dunner1995}}	&	\underline{Ref.~\onlinecite{Sengupta2005}}\\
\hspace{0.8cm} $a$ (\AA)            		& 4.2330(1) 	&	4.215(1)					&	4.260(1)	\\	
\hspace{0.8cm} $c$ (\AA)          			& 10.1683(3)  	&	10.185(2)					&	10.203(1)	\\
\hspace{0.8cm} $c/a$           			& 2.4022(1)  	&	2.416(1)					&	2.395(1)	\\
\hspace{0.8cm} $V_{\rm cell}$  (\AA$^{3}$) 	& 182.20(1)   	&	180.95(12)				&	185.16(11)\\
\hspace{0.8cm} $V_{\rm M}\ ({\rm cm^3/mol}$)	& 54.86	\\
Refinement quality\\
\hspace{0.8cm} $\chi^2$   				& 4.25\\	
\hspace{0.8cm} $R_{\rm p}$ (\%)  			& 3.99\\
\hspace{0.8cm} $R_{\rm wp}$ (\%) 			& 5.48\\
\hline
&${\rm EuCu_2Sb_2}$ \\
Lattice parameters						& \underline{This Work} 	&	\underline{Ref.~\onlinecite{Dunner1995}}	\\
\hspace{0.8cm} $a$ (\AA)            		&  4.4876(2)	&	4.504(1)	\\	
\hspace{0.8cm} $c$ (\AA)          			&  10.7779(5)  &	10.824(2)\\
\hspace{0.8cm} $c/a$           			& 2.4017(2)  	&	2.403(1)	\\
\hspace{0.8cm} $V_{\rm cell}$  (\AA$^{3}$) 	&  217.05(3)   &	219.58(14)\\
\hspace{0.8cm} $V_{\rm M}\ ({\rm cm^3/mol}$)	& 65.35	\\
Refinement quality\\
\hspace{0.8cm} $\chi^2$   & 4.21 \\	
\hspace{0.8cm} $R_{\rm p}$ (\%)  & 4.76 \\
\hspace{0.8cm} $R_{\rm wp}$ (\%) & 6.62 \\

\end{tabular}
\end{ruledtabular}
\end{table}

\begin{table}
\caption{\label{tab:XRD-coord} Atomic coordinates and occupancies (Occ) obtained from the Rietveld refinements of powder XRD data for crushed crystals of EuCu$_2$As$_2$ and EuCu$_2$Sb$_2$.  Also shown are single-crystal data for the As~$z$ parameter from Ref.~\onlinecite{Dunner1995}. The value for this parameter determined in the present work is denoted by ``PW''.}
\begin{ruledtabular}
\begin{tabular}{lcccccc}
   \hspace{0.6cm}  Atom & Wyckoff  & 	Occ	&	 $x$ &	$y$		&	$z$	  	&	$z$ 	\\	
   					& symbol 	& 	(\%)	& 		&			&	(PW)		&	(Ref.~\onlinecite{Dunner1995})	\\
\hline
${\rm EuCu_2As_2}$ \\				
   \hspace{0.8cm}     Eu & $2a$  & 	100	&	 0 		&	0		&   0		&	0	\\
   \hspace{0.8cm}     Cu & $4d$	 & 	98(1)	&	 0 		&	1/2		&	1/4 	  	&	1/4	\\
   \hspace{0.8cm}     As & $4e$  & 	100	&	 0 		&	0 		&	0.3798(2) &	0.3762(1)	\\
${\rm EuCu_2Sb_2}$  \\				
   \hspace{0.8cm}     Eu  & $2c$ & 	100	&	 1/4 	&	1/4		&   0.2381(3)	&	0.2381(1)	 \\
   \hspace{0.8cm}     Cu1 & $2a$ & 101(1)	&	 3/4 	&	1/4		&	0		&	0	 \\
   \hspace{0.8cm}     Cu2 & $2c$ & 82(1)	&	 1/4 	&   1/4 		&	0.6366(9)	&	 0.6365(2)	 \\
   \hspace{0.8cm}     Sb1 & $2b$ & 	100	&	 3/4 	&	1/4		&	1/2 		&	1/2	 \\
   \hspace{0.8cm}     Sb2 & $2c$ & 	100	&	 1/4 	&   1/4 		&	0.8659(4)	&	0.8692(1)	 \\
\end{tabular}
\end{ruledtabular}
\end{table}

The crystallographic and refinement quality parameters obtained from the Rietveld refinements of the XRD data are listed in Tables~\ref{tab:XRD} and \ref{tab:XRD-coord}. While refining the XRD data the occupancies of Eu and As/Sb positions were kept fixed at their stoichiomentric values and the occupation of the Cu site(s) were allowed to vary. The thermal parameters $B$ were also fixed to $B \equiv 0$ as there was no change in the lattice parameters and/or $z$-parameters within the error bars upon varying $B$.  The occupancies are found to be stoichiomentric except for the Cu2 position in \cusb\ which showed 18(1)\% vacancies, for an overall composition ${\rm EuCu_{1.82(1)}Sb_2}$.  These occupancies are in agreement with our EDS results in Sec.~\ref{ExpDetails} which suggested the presence of Cu vacancies in \cusb\ but not in \cuas.  Also shown in Tables~\ref{tab:XRD} and \ref{tab:XRD-coord} for comparison are the lattice parameters and atomic positions for single crystals of both compounds reported in Ref.~\onlinecite{Dunner1995} and the lattice parameters for a polycrystalline sample of ${\rm EuCu_2As_2}$ reported in Ref.~\onlinecite{Sengupta2005}.  

The differences in the lattice parameters and unit cell volumes in Table~\ref{tab:XRD} reported in the three studies of \cuas\ and \cusb\ are outside the respective error bars stated, suggesting that the different samples of \cuas\ and \cusb\ can have different compositions.  This variability evidently arises from variations in the Cu site occupancies.  Even though the unit cell volumes of \cuas\ and \cusb\ differ by almost 20\%, the $c/a$ ratios are nearly identical at $\approx 2.40$.

\section{\label{EuCu2Sb2} Physical Properties of E\lowercase{u}C\lowercase{u}$_{1.82}$S\lowercase{b}$_2$ Crystals}

We begin with the physical properties of \cusbB\ because the interpretation of the magnetic data for this compound is more straightforward than for \cuas.

\subsection{\label{Sec:EuCu2Sb2ChiMH} Magnetization and Magnetic Susceptibility}

\begin{figure}
\includegraphics[width=3in]{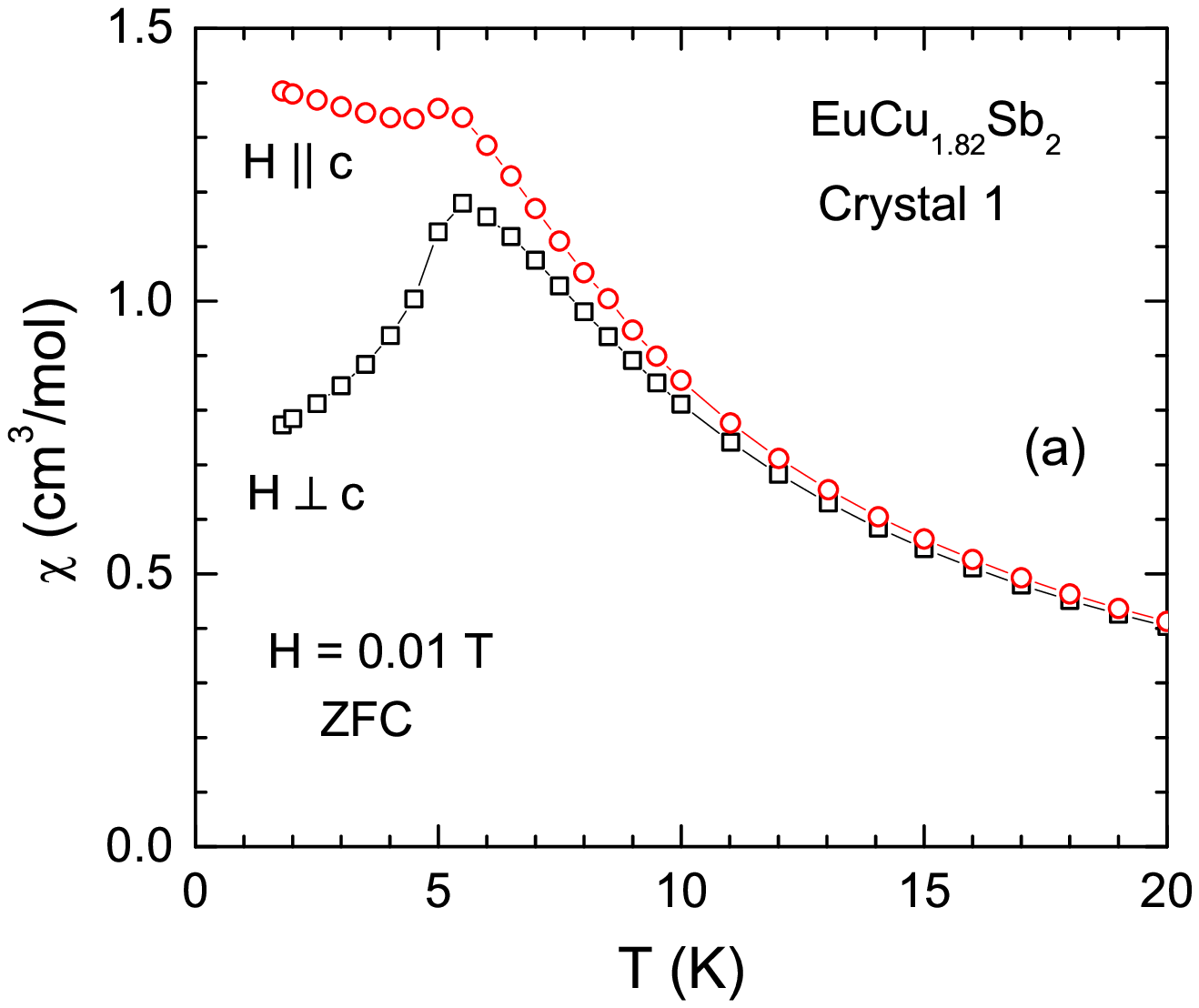}\vspace{0.1in}
\includegraphics[width=3in]{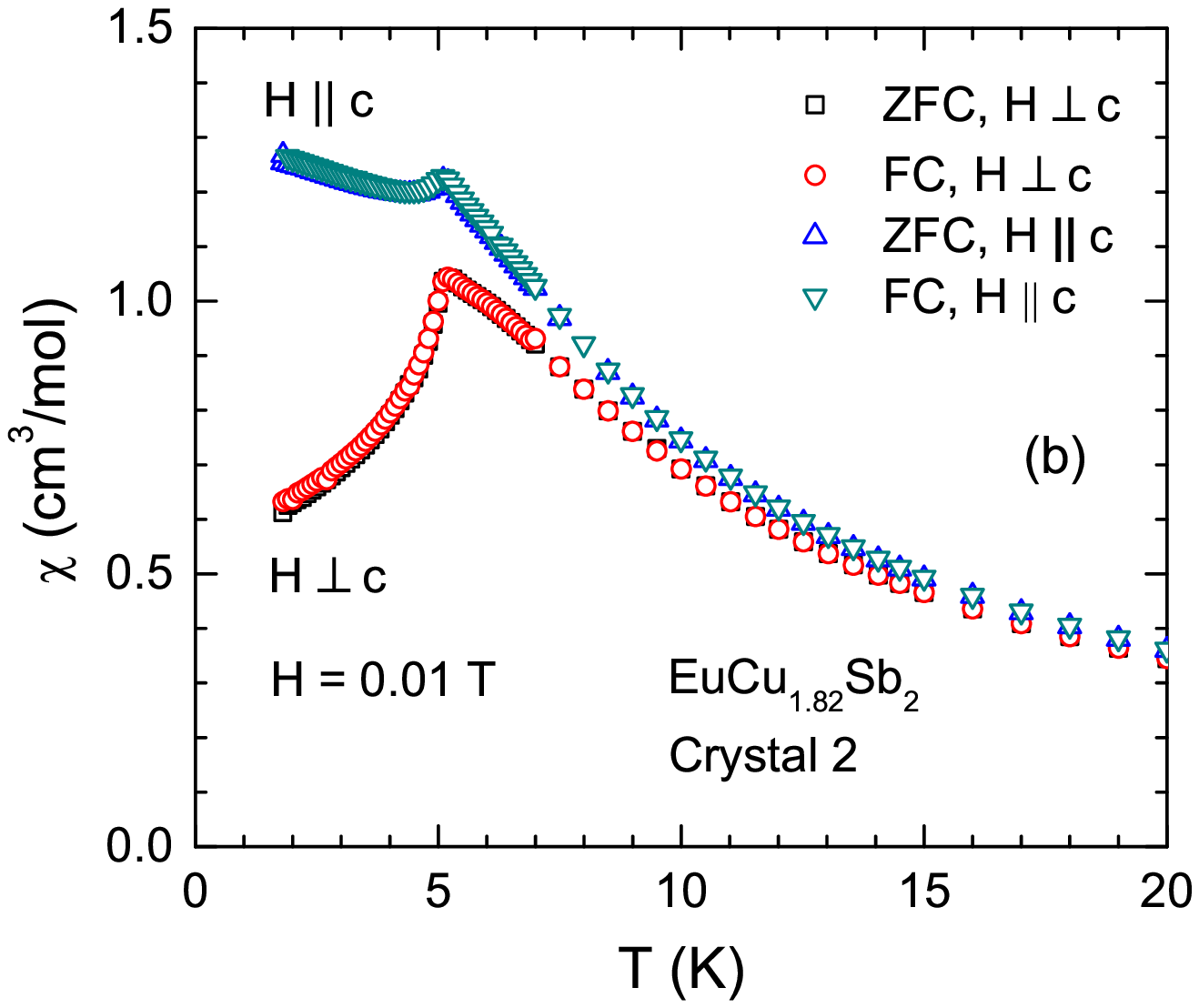}
\caption{(Color online) Zero-field-cooled and field-cooled magnetic susceptibility $\chi$ of \cusbB\ as a function of temperature $T$ in the temperature range 1.8--20~K measured on two different single crystals (crystal~1 and crystal~2) in a magnetic field $H= 0.01$~T applied along the $c$~axis ($\chi_c, H \parallel c$) and in the $ab$~plane ($\chi_{ab}, H \perp  c$).}
\label{fig:MT_EuCu2Sb2_low-H}
\end{figure}

\begin{figure}
\includegraphics[width=3in]{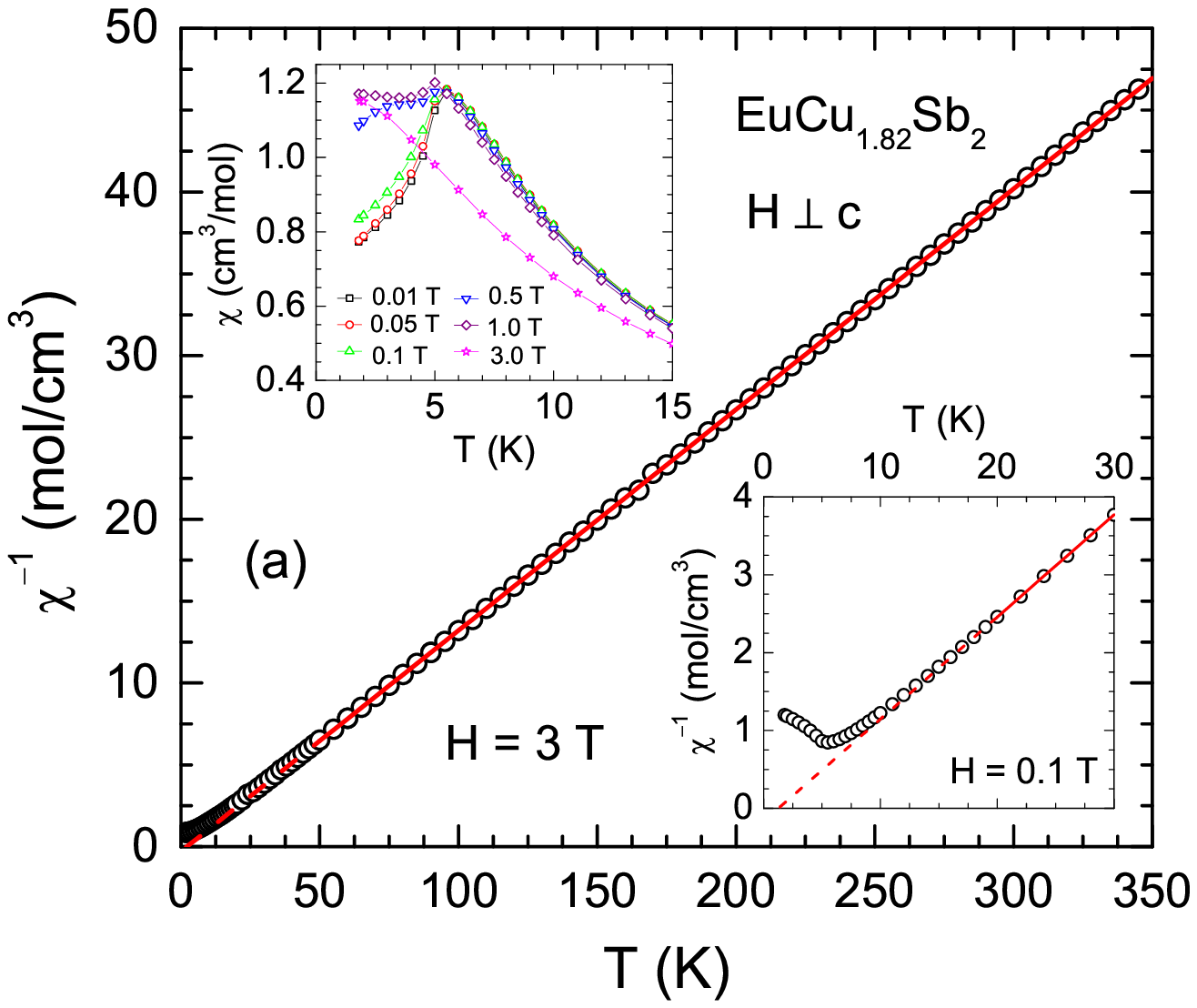}\vspace{0.1in}
\includegraphics[width=3in]{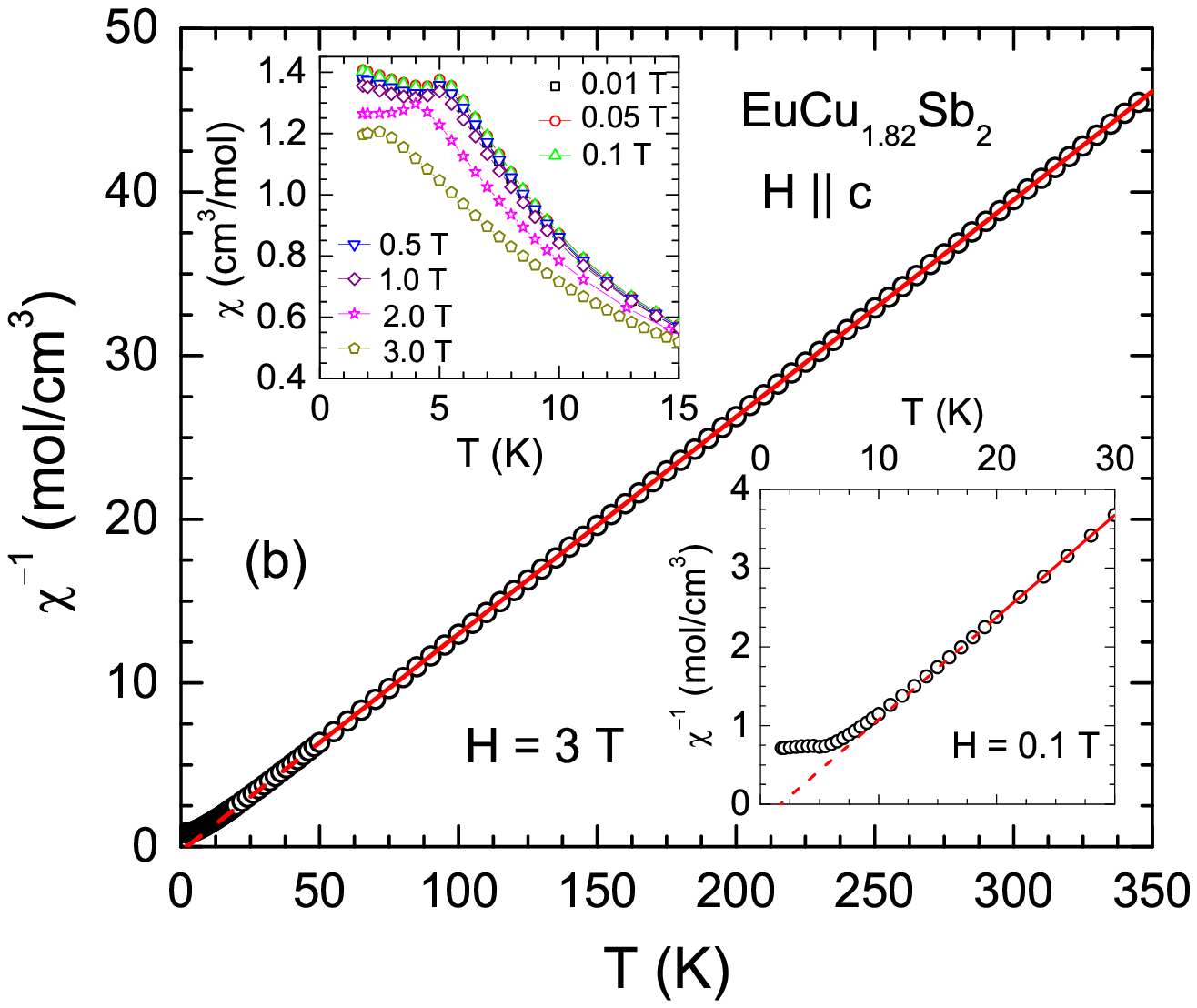}
\caption{(Color online) Zero-field-cooled inverse magnetic susceptibility $\chi^{-1}$ of an \cusbB\ single crystal (crystal~1) as a function of temperature $T$ in the temperature range 1.8--350~K measured in a magnetic field of 3.0~T applied (a) in the $ab$-plane ($\chi_{ab}, H \perp  c$) and, (b) along the $c$-axis ($\chi_c, H \parallel c$). The solid straight red lines are fits of the $\chi^{-1}(T)$ data by the Curie-Weiss law~(\ref{eq:C-W}) in the $T$ range  50~K~$\leq T \leq$~350~K\@. The upper insets in (a) and (b) show $\chi(T)$ at low~$T$ and at different applied fields and the lower insets show fits of the Curie-Weiss law~(\ref{eq:C-W}) to $\chi^{-1}(T)$ data between 20 and 30~K in $H=0.1$~T\@.}
\label{fig:MT_EuCu2Sb2}
\end{figure}

The zero-field-cooled (ZFC) and field-cooled (FC) magnetic susceptibilities $\chi  \equiv M/H$ of an \cusbB\ single crystal as a function of temperature $T$ measured at different $H$ aligned along the $c$~axis ($\chi_c,\ H \parallel c$) and in the $ab$~plane ($\chi_{ab},\  H \perp c$) are shown in Figs.~\ref{fig:MT_EuCu2Sb2_low-H} and \ref{fig:MT_EuCu2Sb2}.  At low $H$ (e.g., at $H = 0.01$~T, Fig.~\ref{fig:MT_EuCu2Sb2_low-H}) sharp anomalies are observed at $T = 5.1$~K in both $\chi_{ab}$ and $\chi_c$ that we identify as the AFM ordering (N\'eel) temperature $T_{\rm N}$. No hysteresis is observed in ZFC and FC $\chi(T)$ data [Fig.~\ref{fig:MT_EuCu2Sb2_low-H}(b)]. An increase in $H$ results in a shift of $T_{\rm N}$ to lower~$T$ (upper insets of Fig.~\ref{fig:MT_EuCu2Sb2}), consistent with expectation for an AFM transition.  The $\chi(T<T_{\rm N})$ in Fig.~\ref{fig:MT_EuCu2Sb2_low-H} is anisotropic with $\chi_{ab} < \chi_c$, indicating that the easy axis lies in the $ab$ plane (collinear AFM ordering) or the easy plane is the $ab$~plane (planar noncollinear AFM ordering).  This easy-plane behavior is common for AFM ordering in the iron arsenide family, e.g., in $A{\rm Fe_2As_2}$ ($A$ = Ca, Sr, Ba, Eu).\cite{Johnston2010,Jiang2009c}  As discussed in Sec.~\ref{Sec:MFA} below, the data in Fig.~\ref{fig:MT_EuCu2Sb2_low-H} are consistent with either a collinear A-type AFM structure with domains with orthogonal easy axes, or a planar noncollinear helix with the helix axis being the $c$~axis.  The A-type AFM candidate structure of \cusbB\ for a single domain is illustrated in Fig.~\ref{fig:structure}(b).  

In the paramagnetic (PM) state the $\chi(T>T_{\rm N})$ data follow the Curie-Weiss behavior in Eq.~(\ref{eq:C-W}).  The linear fits of $\chi^{-1}(T)$ data (shown by straight red lines in Fig.~\ref{fig:MT_EuCu2Sb2}) measured at $H=3$~T in the temperature range 50~K~$\leq T \leq$~350~K by the Curie-Weiss law yield the Curie constants $C$ and Weiss temperatures $\theta_{\rm p}$ listed in Table~\ref{tab:CW}, along with the values of the parameter $f = \theta_{\rm p}/T_{\rm N}$.  Also listed are the effective magnetic moments calculated from the values of $C$ using the relation $\mu_{\rm eff} =  \sqrt{8C}$.  The values of $\mu_{\rm eff}$ are similar to the theoretical value $\mu_{\rm eff} = g\sqrt{S(S+1)}\,\mu_{\rm B} = 7.94\, \mu_{\rm B}$ for a free Eu$^{+2}$ ion with spin $S=7/2$ and spectroscopic splitting factor $g=2$. This indicates that the Eu ions in \cusbB\ are divalent.

\begin{table}
\caption{\label{tab:CW} Antiferromagnetic ordering temperature $T_{\rm N}$ and the parameters obtained from Curie-Weiss fits of the magnetic susceptibility data for ${\rm EuCu_2As_2}$ and \cusbB, where $C$ is the Curie constant and $\theta_{\rm p}$ is the Weiss temperature.  The parameter $f$ is defined as $f = \theta_{\rm p}/T_{\rm N}$.  The effective magnetic moment per Eu atom, $\mu_{\rm eff}$, is obtained from~$C$ according to $\mu_{\rm eff} = \sqrt{8C}$.  The $H=3$~T data were fitted by the Curie-Weiss law from 50 to 350~K, whereas the $H=0.1$~T data were fitted from 20 to 30~K to obtain more accurate estimates of the Weiss temperatures.}
\begin{ruledtabular}
\begin{tabular}{lcccccc}
Compound & Field   & $T_{\rm N}$  & $C$  &  $\theta_{\rm p}$   & $f$ &	$\mu_{\rm eff}$ \\
& direction & (K) & (${\rm \frac{cm^3\,K}{mol}}$) & (K) & &($\frac{\mu_{\rm B}}{\rm Eu}$) \\
\hline
\underline{$H = 3$~T} \\
${\rm EuCu_2As_2}$ & $H \perp c$     & 17.5 &  7.44(1)  & +19.0(2) & 	1.09	&	7.72(1)   \\	
                   & $H \parallel c$ & 17.5 &  7.64(3)  & +17.2(5) &  0.98	&	7.82(2)  \\				
\cusbB\ & $H \perp c$     & 5.1 &  7.41(1)  & +2.2(4) &  0.43	&	7.70(1)   \\	
                   & $H \parallel c$ & 5.1 &  7.54(1)  & +2.1(1) &  0.41	&	7.77(1)  \\
\underline{$H = 0.1$~T} \\
\cusbB\ & $H \perp c$     & 5.1 &  7.62(1)  & +1.26(4) &  0.25	&	7.81(1)   \\	
                   & $H \parallel c$ & 5.1 &  7.71(1)  & +1.69(4) &  0.33	&	7.85(1)  \\
\end{tabular}
\end{ruledtabular}
\end{table}

\begin{figure}
\includegraphics[width=3in]{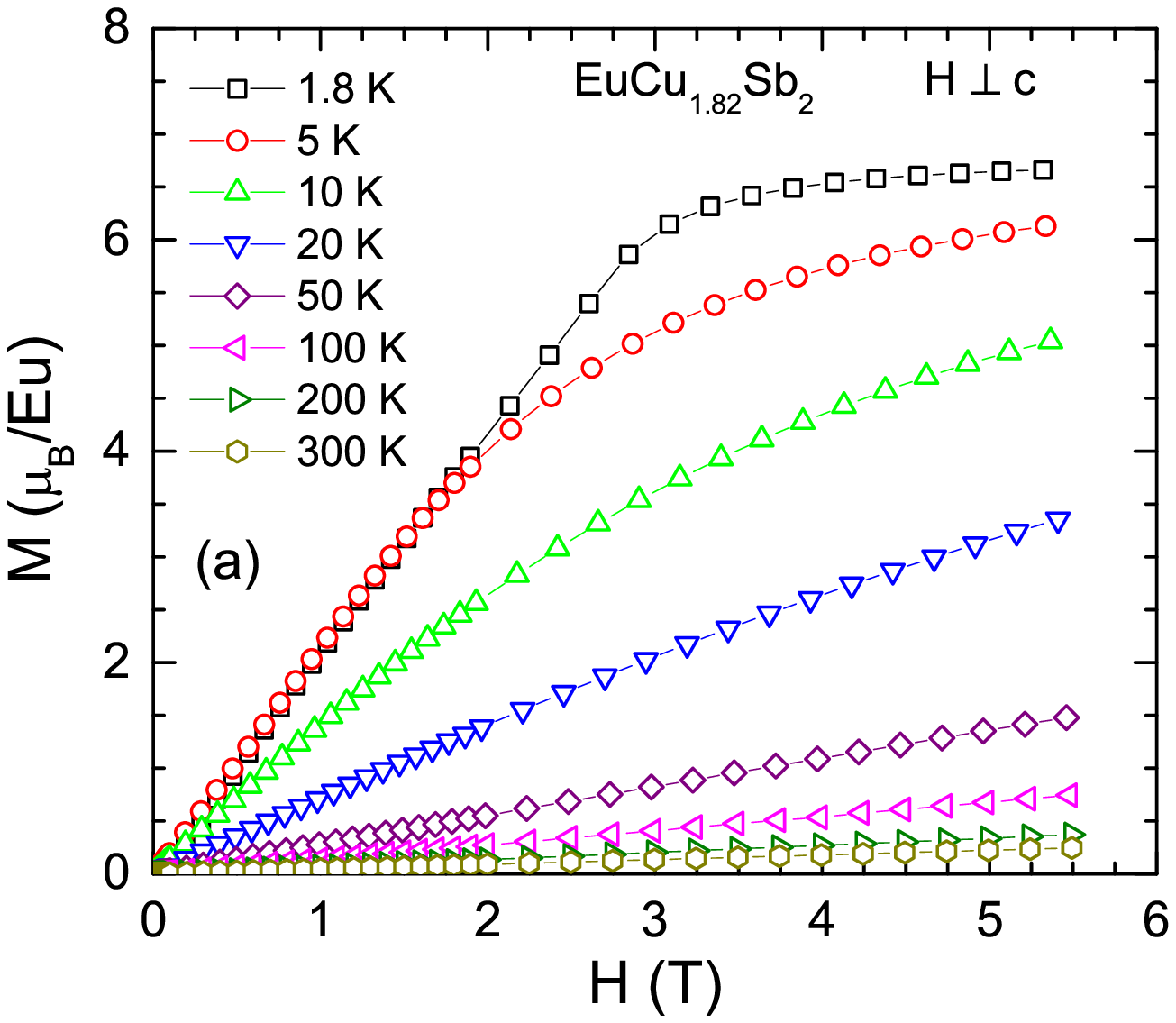}\vspace{0.1in}
\includegraphics[width=3in]{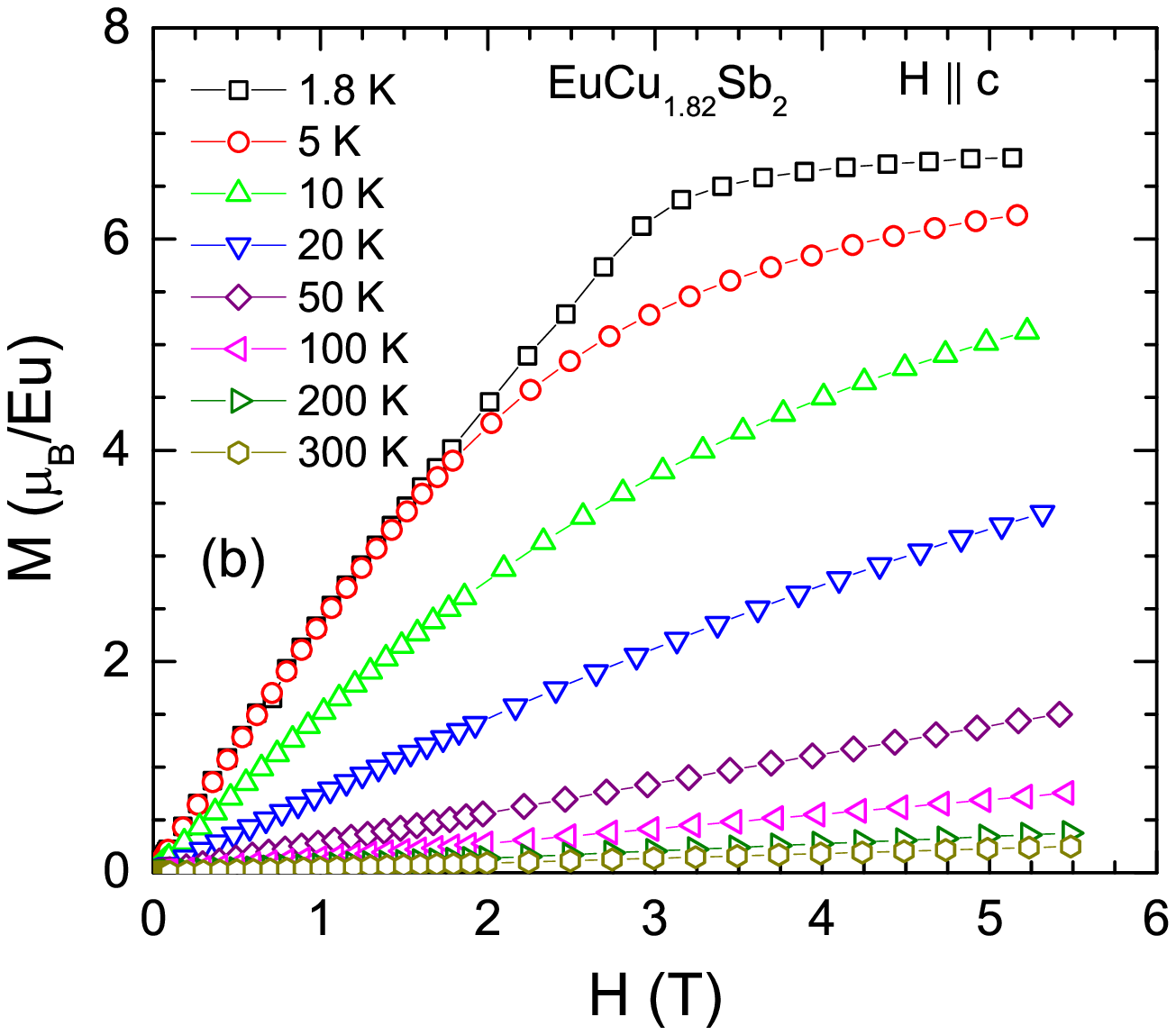}
\caption{(Color online) Isothermal magnetization $M$ of an \cusbB\ single crystal (crystal~1) as a function of internal magnetic field $H$ measured at the indicated temperatures for $H$ (a) in the $ab$~plane ($M_{ab}, H \perp  c$) and (b) along the $c$~axis ($M_c, H \parallel c$).}
\label{fig:MH_EuCu2Sb2}
\end{figure}

\begin{figure}
\includegraphics[width=3in]{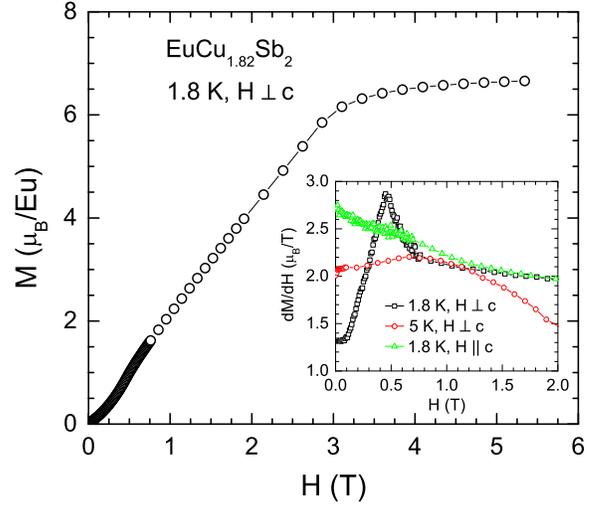}
\caption{(Color online) Isothermal magnetization $M$ of an \cusbB\ single crystal (crystal~2) as a function of internal magnetic field $H$ measured at 1.8~K for $H \perp c$. Inset: The derivative $dM/dH$ versus $H$ at 1.8~K for $H \perp  c$ and $H \parallel c$, and at 5~K for $H\perp c$.}
\label{fig:MH_EuCu2Sb2_2K}
\end{figure}

To further clarify the nature of the AFM structure, $M(H)$ isotherms were obtained at eight temperatures between 1.8 and 300~K for $H$ applied along the $c$~axis ($M_c, H \parallel c$) and in the $ab$~plane ($M_{ab}, H \perp  c$) as shown in Figs.~\ref{fig:MH_EuCu2Sb2} and \ref{fig:MH_EuCu2Sb2_2K}.  Figure~\ref{fig:MH_EuCu2Sb2} shows that at 1.8~K, $M$ increases almost linearly with $H$ up to $H\sim 3$~T (except for a slope change near $H= 0.5$~T for $H \perp  c$), above which $M$ tends to saturate with saturation moments $\mu_{\rm sat}^{ab} = 6.66\,\mu_{\rm B}$ and $\mu_{\rm sat}^{c} = 6.77\,\mu_{\rm B}$ at $H=5.5$~T for $M_{ab}$ and $M_c$, respectively.  These saturation moments are close to the expected $\mu_{\rm sat} = gS\mu_{\rm B}$/Eu $=7\mu_{\rm B}$/Eu assuming $g=2$ and $S=7/2$.  No hysteresis was observed between the increasing and decreasing cycles of $H$ for the $M(H)$ isotherms at 1.8~K (data not shown). At 5~K $\approx T_{\rm N}=5.1$~K, the $M(H)$ curves in Fig.~\ref{fig:MH_EuCu2Sb2} for the two field directions exhibit very similar behaviors, as expected for the PM state.  At temperatures $T \geq 50$~K, $M$ is nearly proportional to~$H$ as expected in the PM regime with $g\mu_{\rm B}H/(k_{\rm B}T) \ll 1$ for $H\leq 5.5$~T\@.

We obtained high-resolution $M(H)$ data at 1.8~K with $H \perp c$ for $H \leq 0.75$~T as shown in Fig.~\ref{fig:MH_EuCu2Sb2_2K}.  One observes a nonlinear behavior of $M(H)$ in this field range.  To study this nonlinearity in detail, shown in the inset of Fig.~\ref{fig:MH_EuCu2Sb2_2K} are plots of $dM/dH$ versus $H$ for $H\perp c$ at 1.8 and 5~K and for $H\parallel c$ at 1.8~K\@. For $H\parallel c$ at 1.8~K, $dM/dH$ versus $H$ is featureless, confirming that the $ab$~plane is the easy plane and the $c$~axis is a hard axis.  The $dM/dH$ versus $H$ data at 5~K for $H\perp c$ is also featureless, as expected for the PM state.  

On the other hand, the  $dM/dH$ versus $H$ data at 1.8~K for $H\perp c$ in the inset of Fig.~\ref{fig:MH_EuCu2Sb2_2K} exhibit two important features.  First $\chi=dM/dH$ is constant for $0 < H \lesssim0.1$~T\@. Second, for $H > 0.1$~T, $dM/dH$ increases and exhibits a pronounced peak at $H\approx 0.5$~T\@. These two features suggest a spin-flop transition that is distributed in field.  In such a transition, the ordered moments flop from an orientation that is not perpendicular to $H$ to a perpendicular orientation.  The field scale over which the magnetization changes sharply in the inset of Fig.~\ref{fig:MH_EuCu2Sb2_2K} ($\lesssim 0.7$~T) at $T=1.8$~K and $H\perp c$ is of the order expected from a spin-flop transition associated with magnetic dipole interaction anisotropy as deduced from the data in Table~\ref{tab:MagDipole0} below, where the ordered moment direction flops from the low-susceptibility direction parallel to the $ab$ plane to the high-susceptibility direction approximately parallel to the $c$~axis (see Fig.~\ref{fig:MT_EuCu2Sb2_low-H}).  At higher fields the ordered moments increasingly point towards the applied field direction until they are parallel to the applied field with parallel moment approximately equal to the saturation moment of $\approx 7~\mu_{\rm B}$/Eu, which happens at the critical field $H_{\rm c}\approx 3$~T as seen in Fig.~\ref{fig:MH_EuCu2Sb2_2K}, which is discussed in quantitative detail in the following section.  For $H>H_{\rm c}$ the system is in the PM phase.

\subsection{\label{Sec:MFA} Molecular Field Theory Analysis of Magnetic Properties}

Because the Eu spin $S=7/2$ is large, the quantum fluctuations associated with finite spin should be small and we expect molecular field theory (MFT) to fit the experimental magnetic susceptibility and heat capacity data rather well.  We use a version of MFT for antiferromagnets developed by one of us for systems of identical crystallographically-equivalent spins interacting by Heisenberg exchange that does not use the concept of magnetic sublattices.\cite{Johnston2011, Johnston2012, Johnston2015}  Instead, the magnetic and thermal properties are calculated solely from the exchange interactions of an arbitrary spin with its neighbors.

Here we discuss two candidates for the magnetic structure.  The first is multiple domains of a collinear A-type AFM with the ordered moments aligned in the $ab$ plane, and the second is a planar noncollinear helical AFM structure with the ordered moments again aligned in the $ab$~plane which is perpendicular to the helix $c$~axis.  A figure showing the helical AFM structure is given in Ref.~\onlinecite{Johnston2012}.  Here we only consider Heisenberg exchange interactions between the spins.  We discuss later how some of the predictions are modified by the presence of magnetic dipole interactions.  

\subsubsection{\label{Sec:TypeAOrder} Magnetic Susceptibility: Collinear Antiferromagnetic Ordering}

In the Weiss MFT for a system of identical crystallographically equivalent spins interacting by Heisenberg exchange with Hamiltonian ${\cal H} = \sum_{<ij>}J_{ij}{\bf S}_i\cdot{\bf S}_j$, where the sum is over distinct pairs of spins interacting with exchange constants $J_{ij}$ and a positive $J_{ij}$ corresponds to an AFM interaction and a negative one to a FM interaction, the magnetic susceptibility $\chi_{\parallel}$ parallel to the easy axis of a collinear AFM at $T\leq T_{\rm N}$ is given by the law of corresponding states\cite{Johnston2012,Johnston2015}
\bse
\begin{equation}
\frac{\chi_{\parallel}(T)}{\chi(T_{\rm N})} = \frac{1-f}{\tau^*(t)-f},
\label{eq:Chi_parallel}
\end{equation}
where 
\be
f = \frac{\theta_{\rm p}}{T_{\rm N}}, \quad t =\frac{T}{T_{\rm N}},\quad \tau^*(t) = \frac{(S+1)t}{3B'_S(y_0)}, \quad y_0 = \frac{3\bar{\mu}_0}{(S+1)t}, 
\ee
the ordered moment versus temperature in zero field is denoted by $\mu_0$, the reduced ordered moment $\bar{\mu}_0 = \mu_0/\mu_{\rm sat} = \mu_0/(gS\mu_{\rm B})$ is determined by solving
\be
\bar{\mu}_0 = B_S(y_0),
\label{Eq:barmu0}
\ee
$B'_S(y_0) = [dB_S(y)/dy]|_{y=y_0}$, and our unconventional definition of the Brillouin function $B_S(y)$ is
\be
B_S(y) = \frac{1}{2S} \left\{(2S+1)\coth\left[(2S+1)\frac{y}{2}\right]-\coth\left(\frac{y}{2}\right)\right\}.
\ee
\ese
Within MFT, the susceptibility in the ordered state of an AFM with the field applied perpendicular to the easy axis of a collinear AFM or to the easy plane of a planar noncollinear AFM is independent of $T$, i.e.,
\be
\chi_\perp(T\leq T_{\rm N}) = \chi(T_{\rm N}).
\label{Eq:ChiPerp}
\ee

In the case of A-type AFM ordering with the ordered moments aligned in the $ab$~plane, there are two equivalent orthogonal directions for the easy axis due to the two equivalent orthogonal $a$ and $b$ axes of the tetragonal unit cell.  This gives rise to two equivalent A-type AFM domains with easy axes in the $ab$~plane that are orthogonal to each other.  The fractional populations of the two domains need not be the same.  Let $x$ be the fractional population of domains with the easy axis perpendicular to the applied field with $H\perp c$.  Then the fractional population of domains with the easy axis parallel to the applied field is $1-x$.  The average (measured) susceptility in the $ab$~plane is then
\bse
\label{Eqs:ChiabA-type}
\be
\chi_{ab\,{\rm ave}} = x\chi_\perp + (1-x)\chi_\parallel(T) = x\chi(T_{\rm N}) + (1-x)\chi_\parallel(T),
\ee
where we used Eq.~(\ref{Eq:ChiPerp}) for $\chi_\perp$.  Since $\chi_\parallel(T = 0) = 0$,\cite{Johnston2012,Johnston2015} the value of $x$ is given by
\be
x = \frac{\chi_{ab\,{\rm ave}}(T=0)}{\chi(T_{\rm N})}.
\ee
\ese

\begin{figure}
\includegraphics[width=3in]{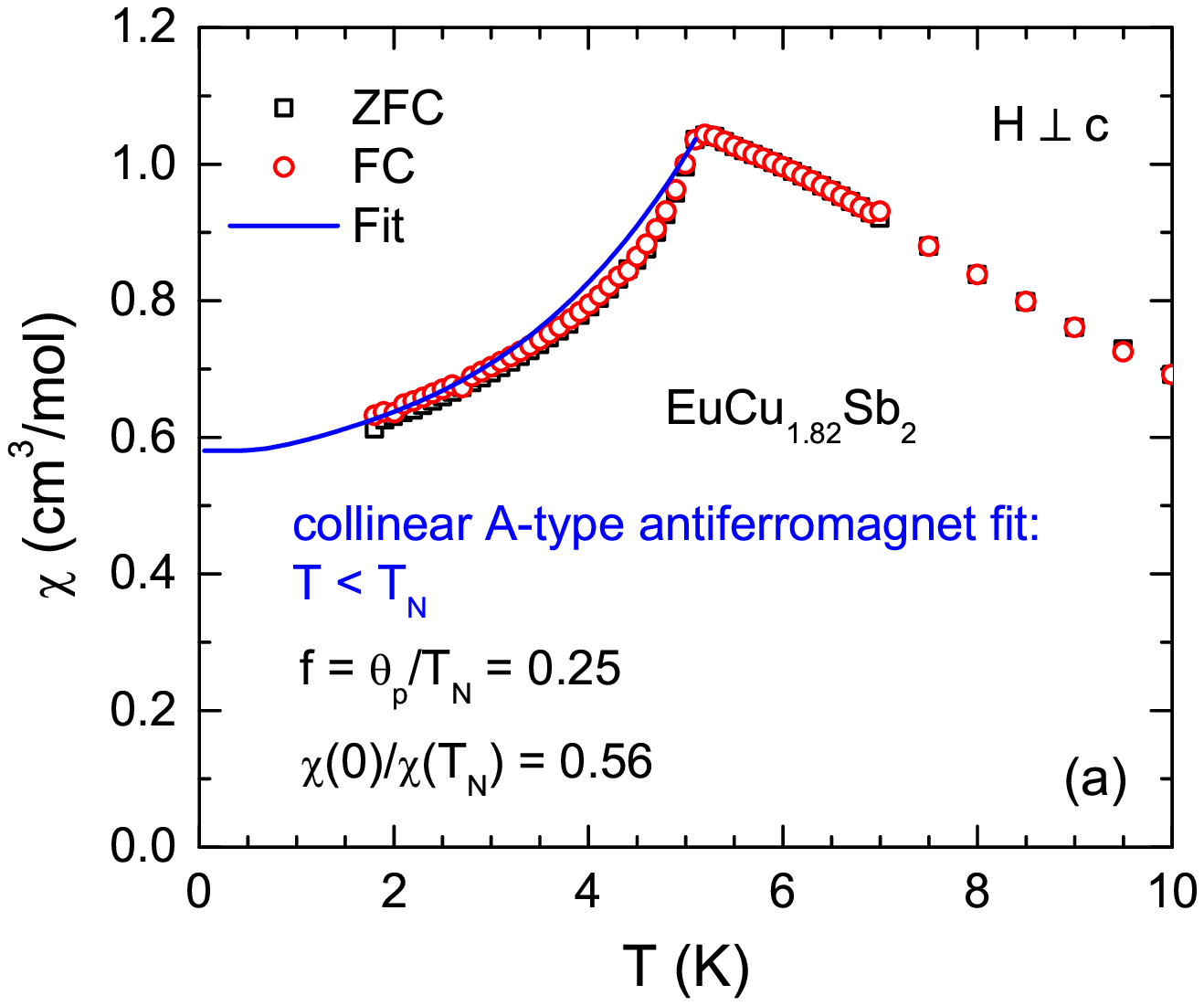}
\includegraphics[width=3in]{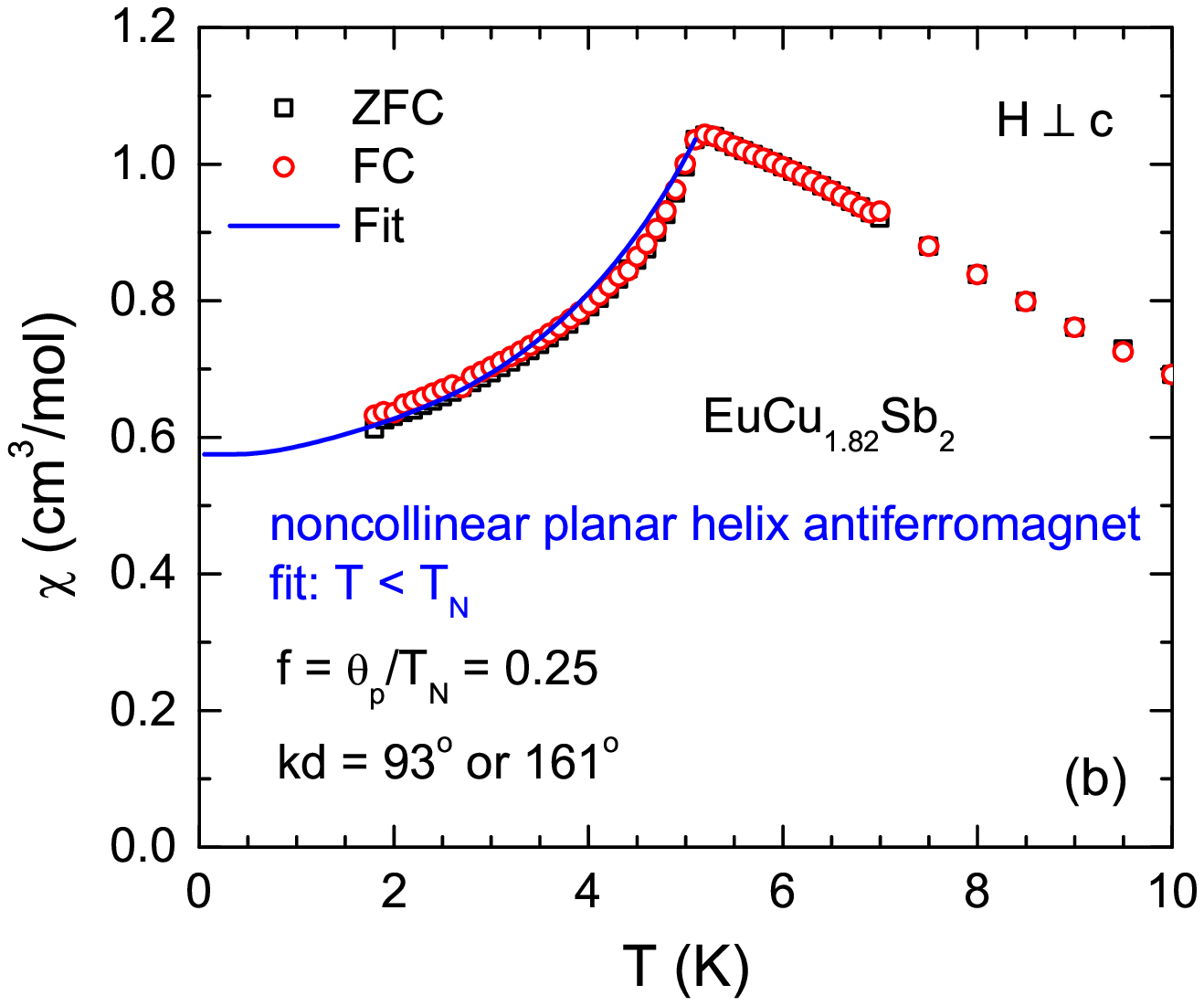}
\caption{(Color online) Fits of the $\chi_{\rm ab}(T\leq T_{\rm N})$ data with $H \perp  c$ (crystal~2) shown in Fig.~\ref{fig:MT_EuCu2Sb2_low-H} for \cusbB\ by MFT for (a) a multi-domain A-type AFM structure and for (b) a $c$-axis helical AFM structure.  The fits are about equally good and hence the fits do not distinguish between the two AFM structures.}
\label{fig:Chi_EuCu2Sb2_A-type}
\end{figure}

From the data in Fig.~\ref{fig:Chi_EuCu2Sb2_A-type} one obtains $x=0.56$.  The additional parameters needed to fit the $\chi_{ab}(T)$ data for $T\leq T_{\rm N}$ by the MFT are $f$, the saturation moment $\mu_{\rm sat} = gS\mu_{\rm B}$ and $\chi(T_{\rm N})$.  For \cusbB\ and $H\perp c$ we use $f=0.25$ from Table~\ref{tab:CW}, $\mu_{\rm sat} = gS\mu_{\rm B} = 7\,\mu_{\rm B}$ using $g=2$ and $S=7/2$, and $\chi(T_{\rm N}) = 1.04~{\rm cm^3/mol}$ from Fig.~\ref{fig:Chi_EuCu2Sb2_A-type}.  The fit of the experimental $\chi_{ab}(T)$ data for $T \leq T_{\rm N}$ by Eqs.~(\ref{eq:Chi_parallel}) and~(\ref{Eqs:ChiabA-type}) is shown by the solid blue curve in Fig.~\ref{fig:Chi_EuCu2Sb2_A-type}(a) with no adjustable parameters.  The experimental data are seen to be well represented by the MFT prediction for collinear A-type AFM ordering with, on average, 56\% of the AFM domains having the easy axis in the $ab$~plane oriented perpendicular to the applied field in the $ab$~plane and 44\% of them having the easy axis parallel to the applied field. 

\subsubsection{Magnetic Susceptibility: Planar Noncollinear Helical Antiferromagnetic Ordering}

The in-plane susceptibility $\chi_{xy}$ for a planar noncollinear helical AFM system is given within MFT by\cite{Johnston2012,Johnston2015}
\bse
\label{Eqs:Chixy}
\begin{equation}
\frac{\chi_{xy}(T \leq T_{\rm N})}{\chi(T_{\rm N})}=  \frac{(1+\tau^*+2f+4B^*)(1-f)/2}{(\tau^*+B^*)(1+B^*)-(f+B^*)^2},
\label{eq:Chi_plane}
\end{equation}
where
\begin{equation}
B^*=  2(1-f)\cos(kd)\,[1+\cos(kd)] - f
\label{eq:Bstar}
\end{equation}
\ese
and $kd$ is turn angle in radians between the ordered moments in adjacent layers along the helix axis. For $T=0$ one obtains the simple result
\be
\frac{\chi_{xy}(T=0)}{\chi(T_{\rm N})} = \frac{1}{2\big[1+2\cos(kd) + 2\cos^2(kd)\big]},
\label{Eq:ChixyT0}
\ee
which only depends on $kd$.  For $0.5 < \chi_{xy}(T=0)/\chi(T_{\rm N}) < 1$ there are two solutions for $kd$ in Eq.~(\ref{Eq:ChixyT0}).\cite{Johnston2012,Johnston2015}

Taking $\chi_{xy}(T=0)/\chi(T_{\rm N}) = 0.56$ from Fig.~\ref{fig:Chi_EuCu2Sb2_A-type} and solving Eq.~(\ref{Eq:ChixyT0}) for $kd$ gives the two solutions $kd = 93^\circ$ and $kd = 161^\circ$ for the turn angle of the helix between adjacent FM-aligned layers along the helix axis in the ordered state of \cusbB\ at $T=0$.  Both solutions correspond to dominant AFM interactions between an ordered moment in one layer and the ordered moments in either of the two adjacent layers because the projection of an ordered moment in one layer on an ordered moment in an adjacent layer is negative.  The predictions of $\chi_{xy}(T\leq T_{\rm N})/\chi(T_{\rm N})$ for the two values of $kd$ are the same and therefore no decision as to which angle is more appropriate is possible from fitting the $\chi(T)$ data alone.  The planar noncollinear helical $\chi(T)$ according to Eqs.~(\ref{Eqs:Chixy}) for $S=7/2$ and $f = 0.25$ with a turn angle $kd=93^\circ$ or 161$^\circ$ is shown in Fig.~\ref{fig:Chi_EuCu2Sb2_A-type}(b).  The $\chi(T\leq T_{\rm N})$ data are described by the helical AFM model and the A-type AFM model equally well.

The magnetic structure for this helical model can be visualized from Fig.~\ref{fig:structure}(b) with the difference that now the angle between the moments of adjacent layers along the $c$~axis is $93^\circ$ or $161^\circ$ as opposed to $180^\circ$ for the A-type AFM model.  However, from a comparison of Figs.~\ref{fig:Chi_EuCu2Sb2_A-type}(a) and~\ref{fig:Chi_EuCu2Sb2_A-type}(b), one sees that when A-type AFM domains with orthogonal easy axes within the $ab$~plane are present in a compound (which make an angle of 90$^\circ$ to each other in the two types of domains), the temperature dependence of the in-plane $\chi_{xy}$ is very similar to that of the corresponding helical AFM with a turn angle of 93$^\circ$ (or 161$^\circ$).

\subsubsection{\label{Sec:MagDipoles} Influence of Magnetic Dipole Interactions}

The response of a magnetic moment in a sample is determined by the local effective magnetic induction it sees.  After crystal shape effects are accounted for, as is consistently done in the $M(H)$ and $\chi(T)$ data presented throughout this paper, one has
\be
B_{{\rm int}\,\alpha\,i}^{\rm local} = H_{0\alpha} + \frac{4\pi}{3}  M_\alpha + B_{{\rm int}\,\alpha\,i}^{\rm near},
\label{Eq:Bintalphai30}
\ee
where the second term is the macroscopic Lorentz field inside a spherical Lorentz cavity, $M_\alpha = \mu/V_{\rm spin}$ is the magnetic moment per unit volume, and the volume per spin is $V_{\rm spin} = V_{\rm cell}/2 = a^2c/2$. The third term $B_{{\rm int}\,\alpha\,i}^{\rm near}$ is the contribution to the local magnetic induction in the $\alpha^{\rm th}$ direction due to the discrete point magnetic dipoles around a central spin within the Lorentz cavity centered on the central spin given by\cite{Lax1952, Rotter2003}
\be
B_{{\rm int}\,\alpha\,i}^{\rm near} = -2\frac{E_i}{\mu} = \frac{\mu\lambda_{{\bf k}\alpha}}{a^3},
\label{Eq:Bi0}
\ee
where the factor of two is necessary in the first equality because $E_i$ is evenly split between the central moment and a neighbor, whereas $B_{{\rm int}\,\alpha\,i}^{\rm near}$ arises only from the neighbor and $\lambda_{{\bf k}\alpha}$ is the eigenvalue of the magnetic dipole interaction tensor $\widehat{{\bf G}}_i({\bf k})$ defined as follows.  The eigenenergies $E_{i}$ are
\bse
\label{Eqs:EiSoln2}
\be
E_{i} = -\epsilon\ \hat{\mu}_i^{\rm T}\widehat{{\bf G}}_i({\bf k})\hat{\mu}_i,
\label{Eq:Ei30}
\ee
where
\be
\epsilon = \frac{\mu^2}{2a^3}
\label{Eq:eps0}
\ee
has dimensions of energy, $a$ is the tetragonal $a$-axis lattice parameter of a simple tetragonal or body-centered tetragonal spin lettice and
\be
\widehat{{\bf G}}_i({\bf k}) = \sum_{j\neq i}\frac{1}{(r_{ji}/a)^5}\bigg(3\frac{{\bf r}_{ji}{\bf r}_{ji}}{a^2} - \frac{r_{ji}^2}{a^2}{\bf 1}\bigg)e^{i{\bf k}\cdot{\bf r}_{ji}}\label{Eq:Gk2}
\ee
is a dimensionless interaction tensor for collinear magnetic ordering, ${\bf r}_{i}$ is the position in Cartesian coordinates of a central spin~$i$ at which the net magnetic induction due to spins at positions ${\bf r}_{j}$ is calculated, ${\bf r}_{ji} = {\bf r}_{j} - {\bf r}_{i},\ r_{ji}=|{\bf r}_{ji}|$ and {\bf k} is the magnetic wave vector.  Labeling the eigenvalues of $\widehat{{\bf G}}_i({\bf k})$ as $\lambda_{{\bf k}\alpha}$, Eq.~(\ref{Eq:Ei30}) gives the eigenenergies
\be
E_{i\alpha} = -\epsilon\ \lambda_{{\bf k}\alpha},
\label{Eq:Eilambda}
\ee
\ese
and the eigenvectors $\hat{\mu}$ are the ordered moment axes of the collinear magnetic structure.  For FM alignment, which can include either an ordered FM structure or an induced alignment due to an applied magnetic field, one has {\bf k} = (0,0,0) and for AFM wave vectors the components of ${\bf k}$ are expressed in conventional primitive-tetragonal reciprocal lattice units (rlu) $2\pi/a$ and $2\pi/c$.

\begin{table*}
\caption{\label{tab:MagDipole0} Eigenvalues $\lambda_{{\bf k}\alpha}$ and eigenvectors $\hat{\mu} = [\mu_x,\mu_y,\mu_z]$ of the magnetic dipole interaction tensor $\widehat{{\bf G}}_i({\bf k})$ in Eq.~(\ref{Eq:Gk2}) for $c/a=2.40$ and six values of the magnetic wave vector {\bf k} for collinear magnetic order.  The accuracy of the $\lambda_{{\bf k}\alpha}$ values is estimated to be $\pm 0.001$. The $E_i$ values for \cuas\ and \cusbB\ were obtained from Eq.~(\ref{Eq:Eilambda}) using the $\lambda_{{\bf k}\alpha}$ values in the second column and the $\epsilon$ values in Eqs.~(\ref{Eqs:epsilons0}) which assume that the ordered moment magnitude is $\mu=7\,\mu_{\rm B}$/Eu.  Also shown are local magnetic induction values $B_{{\rm int}\,\alpha\,i}^{\rm near} = -2E_i/(7\mu_{\rm B})$ obtained from Eq.~(\ref{Eq:Bi0}). }
\begin{ruledtabular}
\begin{tabular}{crccccccc}	
{\bf k} & $\lambda_{{\bf k}\alpha}$  	& $\hat{\mu}$ 					& $-E_i$			& $-E_i/k_{\rm B}$ 	&	$B_{{\rm int}\,\alpha\,i}^{\rm near}$		& $-E_i$		& $-E_i/k_{\rm B}$	&	$B_{{\rm int}\,\alpha\,i}^{\rm near}$		\\
				&				&							&\cuas			& \cuas			&	\cuas		& \cusbB		& \cusbB 			&	\cusbB		\\
				&				&							& ($\mu$eV) 		& (K) 			&	(G)		& ($\mu$eV) 		& (K)			&	(G)		\\
\hline
$\left(\frac{1}{2},\ 0,\ 0\right)$ & 5.100 & [0,\ 1,\ 0] 				&	88.5		&	1.026			&	4366		&	74.2		&	0.861		&	3664		\\
 							& 0.977 & [0,\ 0,\ 1] 				&	17.0		&	0.197			&			&	14.2		&	0.165	\\
 							& $-6.077$ & [1,\ 0,\ 0] 			&	$-105.4$	&	$-1.223$			&			&	$-88.5$	&	$-1.026$	\\
$\left(\frac{1}{2},\ 0,\ \frac{1}{2}\right)$ & 5.100 & [0,\ 1,\ 0] 		&	88.4		&	1.026			&	4364		&	74.2		&	0.861 		&	3664		\\
 					& 1.328 & [0.2366,0,$-0.9716$] 				&	23.0		&	0.267			&			&	19.3		&	0.224	\\
 				& $-6.428$ & [$-0.9716$,0,$-0.2366$] 				&	$-111.5$		&	$-1.293$		&			&	$-93.6$	&	$-1.085$	\\
$\left(0,\ 0,\ \frac{1}{2}\right)$ & 4.517 & [1,\ 0,\ 0] 				&	78.3		&	0.909			&	3866		&	65.7		&	0.763		&	3244		\\
							& 4.517 & [0,\ 1,\ 0] 				&	78.3		&	0.909			&			&	65.7		&	0.763	\\
							& $-9.035$ & [0,\ 0,\ 1]				&	$-156.7$	&	$-1.818$			&			&	$-131.5$	&	$-1.525$	\\
(0,\ 0,\ 1) 						& 4.439 & [1,\ 0,\ 0] 			&	77.0		&	0.893			&	3798		&	64.6		&	0.749		&	3188		\\
\{A-type AFM\}							& 4.439 & [0,\ 1,\ 0] 		&	77.0		&	0.893			&			&	64.6		&	0.749	\\
 								& $-8.877$ & [0,\ 0,\ 1]			&	$-153.9$	&	$-1.786$			&			&	$-129.2$	&	$-1.499$	\\
$\left(\frac{1}{2},\ \frac{1}{2},\ 0\right)$ & 2.652 & [0,\ 0,\ 1] 		&	46.0		&	0.534			&	2270		&	38.6		&	0.448		&	1906		\\
& $-0.789$ & $\left[\frac{1}{\sqrt{2}}, -\frac{1}{\sqrt{2}},0\right]$ 	&	$-13.7$	&	$-0.159$			&			&	$-11.5$	&	$-0.133$	\\
 & $-1.863$ & $\left[\frac{1}{\sqrt{2}},\frac{1}{\sqrt{2}},0\right]$		&	$-32.3$	&	$-0.375$			&			&	$-27.1$	&	$-0.315$	\\
$\left(\frac{1}{2},\ \frac{1}{2},\ \frac{1}{2}\right)$ & 2.640 & [0,\ 0,\ 1] &	45.8		&	0.5311			&	2260		&	38.4		&	0.446		&	1896		\\
\{G-type or	& $-1.320$ & $\left[1,0,0\right]$ 						&	$-22.9$	&	$-0.266$		 	&			&	$-19.2$	&	$-0.223$	\\
N\'eel-type AFM\} & $-1.320$ & $\left[0,1,0\right]$					&	$-22.9$	&	$-0.266$			&			&	$-19.2$	&	$-0.223$	\\
(0,\ 0,\ 0) 						& 1.105 & [1,\ 0,\ 0] 			&	19.2		&	0.222			&	946		&	16.1		&	0.187		&	794		\\
\{ferromagnetic					& 1.105 & [0,\ 1,\ 0] 			&	19.2		&	0.222			&			&	16.1		&	0.187	\\
alignment\}						& $-2.210$ & [0,\ 0,\ 1]			&	$-38.3$	&	$-0.445$			&			&	$-32.2$	&	$-0.373$	\\
\end{tabular}
\end{ruledtabular}
\end{table*}

Shown in Table~\ref{tab:MagDipole0} are the eigenvalues $\lambda_{{\bf k}\alpha}$ and eigenvectors $\hat{\mu}$ of $\widehat{{\bf G}}_i({\bf k})$ calculated by direct lattice summation for a variety of moment configurations for body-centered-tetragonal spin lattices with $c/a=2.40$ as found for the Eu positions in \cuas, \cusbB\ and for the Gd positions in ${\rm GdAu_2Si_2}$.\cite{Rotter2003}  For a moment $\mu = gS\mu_{\rm B}$ with $g = 2$ and $S = 7/2$ for Eu$^{+2}$ or Gd$^{+2}$ and the $a$-axis lattice parameters for \cuas\ and \cusbB\ in Table~\ref{tab:XRD}, Eq.~(\ref{Eq:eps0}) gives
\bse
\label{Eqs:epsilons0}
\bea
\epsilon &=& 17.34~\mu{\rm eV} = 0.2012~{\rm K}\quad ({\rm EuCu_2As_2}),\\*
\epsilon &=& 14.55~\mu{\rm eV} = 0.1688~{\rm K}\quad ({\rm EuCu_{1.82}Sb_2}).
\eea
\ese
Using these values, and assuming stoichiomentric formulas \cuas\ and \cusb, the values of $E_{i\alpha}$ for each of the {\bf k} values shown were calculated from Eq.~(\ref{Eq:Ei30}) and are listed in Table~\ref{tab:MagDipole0} in units of both $\mu$eV and~K\@.  The corresponding values of $B_{{\rm int}\,\alpha\,i}^{\rm near}$ obtained using Eq.~(\ref{Eq:Bi0}) are also shown for the lowest-energy eigenvector for each~{\bf k}.   Our eigenvectors for ${\bf k} = \left(\frac{1}{2},0,\frac{1}{2}\right)$ agree with those calculated in Ref.~\onlinecite{Rotter2003}, but our $E_{i\alpha}$ values are about a factor of 5--6 larger in magnitude than in Ref.~\onlinecite{Rotter2003} for unknown reasons.

One sees from Table~\ref{tab:MagDipole0} that the magnetic dipole interaction favors either ${\bf k} = \left(\frac{1}{2},0,0\right)$~rlu or ${\bf k} = \left(\frac{1}{2},0,\frac{1}{2}\right)$~rlu, neither of which corresponds to the AFM structures postulated above for \cusbB.  However, the ordered moment direction in both cases is along the $b$~axis, in agreement with our measurements to be in the $ab$~plane.  The FM structure has the highest energy among the {\bf k} values listed in the table.

For the collinear A-type AFM structure possibibility with {\bf k} = (0,0,1)~rlu discussed in Sec.~\ref{Sec:TypeAOrder}, the lowest-energy eigenvector for this {\bf k} value is also in the $ab$~plane, consistent with the susceptibility data and fit in Figs.~\ref{fig:MT_EuCu2Sb2_low-H} and~\ref{fig:Chi_EuCu2Sb2_A-type}, respectively.  On the other hand, this {\bf k} does not have the lowest ordering energy among the {\bf k} values listed, which means that Heisenberg exchange interactions instead of dipole interactions determine the AFM structure, but with the dipole interaction determining or at least contributing to the easy axis for the ordering.  Of our two postulated AFM structures, neutron diffraction measurements have determined that \cusb\ orders in the A-type AFM structure with {\bf k} = (0,0,1) and the easy axis indeed lies in the $ab$~plane as predicted for the magnetic dipole interaction.\cite{Ryan2015}

Using MFT, the N\'eel temperature $T_{\rm NA}$ for A-type AFM ordering and $\hat{\mu} = (1,0,0)$ arising from the anisotropic magnetic dipole interaction is\cite{Johnston2011,Johnston2012,Johnston2015}
\be
T_{\rm NA} = \frac{g^2S(S+1)\mu_{\rm B}^2\lambda_{{\bf k}\alpha}}{3a^3k_{\rm B}}.
\ee
Setting $g=2$, $S = 7/2$, $\lambda_{{\bf k}\alpha} = 4.439$ from Table~\ref{tab:MagDipole0} and $a = 4.4876$~\AA\ for \cusbB\ from Table~\ref{tab:XRD}, one obtains
\be
T_{\rm NA} = 0.64~{\rm K}\qquad ({\rm EuCu_{1.82}Sb_2}).
\label{Eq:TNAtext}
\ee
Because the local fields due to the exchange and dipolar interactions are additive in their contributions to $T_{\rm N}$ within MFT, this is the amount by which $T_{\rm N}$ increases due to the magnetic dipole interaction.  Since the measurements give $T_{\rm N} = 5.1$~K, this is only a 13\% effect, which indicates that the measured $T_{\rm N}$ is mainly due to exchange interactions.

In the PM state above the N\'eel temperature, all moments are aligned in the same direction $\alpha$ as the applied magnetic field $H_{0\alpha}$ [the magnetic wave vector is {\bf k} = (0,0,0)].  Then in the limit of low field and after correction for shape effects leading to a demagnetizing field, MFT for dipolar interactions yields the Curie-Weiss law
\be
\chi = \frac{C_1}{T - \theta_{\rm pA}},
\ee
where the single-spin Curie constant\cite{Kittel2005} $C_1$ and the Weiss temperature $\theta_{\rm pA}$ for a bct spin lattice are
\bse
\label{Eq:thetap0}
\be
C_1 = \frac{g^2S(S+1)\mu_{\rm B}^2}{3k_{\rm B}},
\ee
\be
\theta_{\rm pA} = \frac{C_1}{a^3}\bigg(\frac{8\pi }{3c/a} + \lambda_{(0,0,0)\alpha}\bigg).
\ee
\ese
Using $g=2$, $S=7/2$, $a=4.4876$~\AA\ and $c/a=2.40$ from Table~\ref{tab:XRD}, and $\lambda_{(0,0,0)[1,0,0]} = 1.105$ and $\lambda_{(0,0,0)[0,0,1]} = -2.210$ from Table~\ref{tab:MagDipole0}, Eqs.~(\ref{Eq:thetap0}) give
\bse
\bea
\theta_{\rm pA} &=& 0.59~{\rm K}\qquad (H\perp c),\\*
\theta_{\rm pA} &=& 0.35~{\rm K}\qquad (H\parallel c).
\label{thetapA0}
\eea
\ese
The contributions of the magnetic dipole interaction to $\theta_{\rm pA}$ for $H\perp c$ and $H\parallel c$ are both positive and hence ferromagneticlike.

\subsubsection{Exchange Constants and Exchange Field}

One can estimate the exchange interactions within MFT\@.   In this discussion we correct for the contributions of magnetic dipole interactions to the observed values of $T_{\rm N}$ and $\theta_{\rm p}$.  According to Eqs.~(\ref{Eq:TNAtext}) and~(\ref{thetapA0}), the $T_{\rm N}$ is increased by 0.64~K, $\theta_{\rm p}(H\perp c)$ is increased by 0.59~K and $\theta_{\rm p}(H\parallel c)$ is increased by 0.35~K due to the magnetic dipole interactions.  Subtracting these values from the observed $T_{\rm N}=5.1$~K, $\theta_{\rm p}(H\perp c) = 1.26$~K and $\theta_{\rm p}(H\parallel c) = 1.70$~K in Table~\ref{tab:CW} gives the contributions due to the exchange interactions $J_{ij}$ as
\bse
\be
T_{{\rm N}J} = 4.5~{\rm K},
\ee
\be
\theta_{{\rm p}J}(H\perp c) = 0.67~{\rm K},\quad \theta_{{\rm p}J}(H\parallel c) = 1.35~{\rm K}.
\ee
\ese
Thus there appears to be another source of anisotropy present in addition to the magnetic dipole interaction.  Taking the spherical average of these two values as an approximation gives
\be
\theta_{{\rm p}J}=0.90~{\rm K}.
\ee
The value $f_J$ arising from the exchange interactions as
\be
f_J = \frac{\theta_{{\rm p}J}}{T_{{\rm N}J}} = 0.20.
\ee

In MFT $\theta_{{\rm p}J}$ and $f_J$ are related to the exchange interactions by\cite{Johnston2012,Johnston2015}
\begin{equation}
T_{{\rm N}J} = -\frac{S(S+1)}{3k_{\rm B}} \sum_j J_{ij} \cos\phi_{ji},
\label{eq:TN}
\end{equation}
\begin{equation}
\theta_{{\rm p}J} = -\frac{S(S+1)}{3k_{\rm B}} \sum_j J_{ij},
\label{eq:thetap}
\end{equation}
and
\begin{equation}
f_J = \frac{\sum_j J_{ij}} {\sum_j J_{ij} \cos\phi_{ji}},
\label{eq:f}
\end{equation}
where the sums are over all neighbors~$j$ with which a central spin~$i$ interacts with respective exchange constant $J_{ij}$ and $\phi_{ji}$ is the angle between ordered moments $\vec{\mu}_j$ and $\vec{\mu}_i$ in the magnetically-ordered state.  For the A-type AFM structure in Fig.~\ref{fig:structure}(b) with a fourfold in-plane nearest-neighbor interaction $J_1$ and twofold interlayer interaction $J_c$ these expressions become
\begin{equation}
T_{{\rm N}J} = -\frac{S(S+1)}{3k_{\rm B}} (4J_1 -8 J_c),
\label{eq:J-TN}
\end{equation}
\begin{equation}
\theta_{{\rm p}J} = -\frac{S(S+1)}{3k_{\rm B}} (4J_1 + 8 J_c)
\label{eq:J-thetap}
\end{equation}
and
\begin{equation}
f_J = \frac{\theta_{{\rm p}J}}{T_{{\rm N}J}} = \frac{J_1 + 2J_c}{J_1 - 2J_c}.
\label{eq:J-f}
\end{equation}
Using $f_J=0.20$, $\theta_{{\rm p}J} = 0.90$~K and $S=7/2$, from Eqs.~(\ref{eq:J-thetap}) and (\ref{eq:J-f}) we obtain
\be
\frac{J_1}{k_{\rm B}} = -0.129~{\rm K},\qquad \frac{J_c}{k_{\rm B}} = 0.043~{\rm K}.
\ee
Thus $J_1$ is FM and $J_c$ is AFM, consistent with the A-type AFM structure in Fig.~\ref{fig:structure}(b).

The exchange field $H_{\rm exch0}$ can be estimated from the values of $J_1 $ and $J_c $ using the relation \cite{Johnston2012,Johnston2015}
\begin{equation}
H_{{\rm exch}\,i} = -\frac{1}{g^2\mu_{\rm B}^2}\sum_j J_{ij} \mu_j \cos\phi_{ji},
\label{eq:Hexch}
\end{equation}
which for the present case is
\begin{equation}
H_{\rm exch0}(T=0) = -\frac{\mu_0}{g^2\mu_{\rm B}^2}[4J_1 - 2J_c],
\label{eq:J-Hexch}
\end{equation}
where $g=2$ and $\mu_0 = 7~\mu_{\rm B}$.  This gives $H_{\rm exch0}(T=0) = 22.4$~kOe\@.  This is about a factor of 7 larger than the dipolar magnetic induction of 3188~G for ${\bf k} = (0,0,1)$ in Table~\ref{tab:MagDipole0}, as expected since the exchange interactions mainly determine $T_{\rm N}$ as noted above.

\subsubsection{High Magnetic Fields Perpendicular to the Easy Axis or Plane of the Antiferromagnetic Structure}

\begin{figure}
\includegraphics[width=3in]{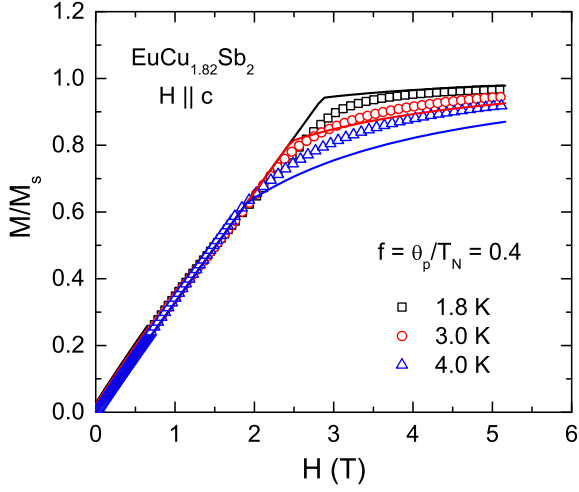}
\caption{(Color online) Magnetization $M$ for crystal~2 of \cusbB, normalized by its saturation value $M_{\rm s}=NgS\mu_{\rm B}$, versus internal magnetic field $H$ for $H \parallel c$ for several temperatures $T < T_{\rm N}=5.1$~K\@.  This field direction is perpendicular to the easy axes or plane of the two candidate AFM structures considered in this paper.  Also shown as solid curves are fits of the data by MFT at the respective temperatures according to Eqs.~(\ref{eq:MH-MFT}).}
\label{fig:MH_EuCu2Sb2_MFT}
\end{figure}

The version of MFT considered in this paper\cite{Johnston2012,Johnston2015} does not utilize the concept of magnetic sublattices, so the thermal and magnetic behavior of each ordered moment in $H=0$ is the same and only depends on its interactions with its neighbors.  Because the response of each ordered moment to a field applied perpendicular to the easy axis of a collinear AFM or to the easy plane of a planar noncollinear AFM is the same since the former structure is a special case of the latter, the same MFT can be applied to obtain a law of corresponding states for a given~$S$ for the response of a given moment to an applied perpendicular field for both types of structures.  We will use these MFT results\cite{Johnston2015} to describe the high-field perpendicular magnetization of \cusbB\ at $T\leq T_{\rm N}$\@.

In MFT, the initial slope of the magnetic moment per spin $\mu_\perp$ of a collinear or planar noncollinear AFM versus perpendicular magnetic field~$H$ is independent of $T$ for $0 \leq T \leq T_{\rm N}$ and the $M(H)$ is given by
\be
\mu_\perp = \chi_\perp H,\qquad \chi_\perp = \chi(T_{\rm N}),
\label{Eq:MperpTleqTN}
\ee
where $\chi(T_{\rm N})$ is the low-field single-spin susceptibility at $T=T_{\rm N}$.  The perpendicular critical field $H_{\rm c\perp}$ is the perpendicular field at which a collinear or planar noncollinear AFM undergoes a second-order transition from the (canted) AFM state to the paramagnetic (PM) state.  As long as $H<H_{\rm c\perp}$, the ordered moment magnitude is independent of field even though the ordered moments are progressively tilting towards the field with increasing field, and the reduced ordered moment is therefore given in this field range by the expression for the zero-field reduced ordered moment $\bar{\mu}_0$ in Eq.~(\ref{Eq:barmu0}).  The critical field is the field at which the angle between the ordered moments and the perpendicular field becomes equal to zero and hence Eq.~(\ref{Eq:MperpTleqTN}) gives
\be
H_{\rm c\perp}(T=0) = \frac{N_{\rm A}\mu_0}{\chi(T_{\rm N})},
\ee
where $N_{\rm A}$ is Avogadro's number, $\mu_0=gS\mu_{\rm B}$ and $\chi(T_{\rm N})$ is here expressed per mole of spins. For $H \geq H_{\rm c\perp}$ the system is in the PM state and hence the magnetic moments in the direction of the field are considered to be purely field-induced moments, as opposed to ordered moments present in the AFM state.   Taking $g=2$ and $S=7/2$ appropriate to Eu$^{+2}$ spins and $\chi(T_{\rm N}) = 1.3~{\rm cm^3/(mol~Eu)}$ from Fig.~\ref{fig:MT_EuCu2Sb2_low-H}(b) gives $H_{\rm c\perp}(T=0)\sim 26$~kOe, similar to the value of $\sim 30$~kOe obtained from Fig.~\ref{fig:MH_EuCu2Sb2}.

It is convenient to define the reduced parameters
\be
h = \frac{g\mu_{\rm B}H}{k_{\rm B}T_{\rm N}},\qquad h_{\rm c\perp} = \frac{g\mu_{\rm B}H_{\rm c\perp}}{k_{\rm B}T_{\rm N}}, \qquad \bar{\mu}_\perp  = \frac{\mu_\perp}{\mu_{\rm sat}},
\ee
where $\mu_{\rm sat} = gS\mu_{\rm B}$.  Then the $H$ dependence of the ordered plus induced moment per spin $\mu_\perp$ for fields applied perpendicular to the easy axis (collinear AFM) or easy plane (planar noncollinear AFM) of a Heisenberg AFM is described by the same law of corresponding states for a given~$S$, given by\cite{Johnston2015}
\begin{equation}
\begin{split}
\bar{\mu}_\perp & = \frac{S+1}{3(1-f)}h  \hspace{2.4cm} (h \leq h_{\rm c\perp}), \\
\bar{\mu}_\perp & = B_S \left[ \frac{3f \bar{\mu}_\perp }{(S+1)t} +\frac{h}{t} \right] \hspace{1cm} (h \geq h_{\rm c\perp}),
\end{split}
\label{eq:MH-MFT}
\end{equation}
where $\bar{\mu}_\perp  = M/M_{\rm s}$,  $M=N\mu_\perp$, $M_{\rm s}=N\mu_{\rm sat}$, $N$ is the number of spins in the system and $t=T/T_{\rm N}$.

The normalized $M(H)/M_{\rm s}$ data for \cusbB\ crystal~2 at $T = 1.8$, 3.0 and 4.0~K are shown for $H\parallel c$ by the open symbols in Fig.~\ref{fig:MH_EuCu2Sb2_MFT}. Also shown are the predictions of MFT for $M(H)/M_{\rm s}$ using Eqs.~(\ref{eq:MH-MFT}) and the value $f=0.4$ for $H\parallel c$ with $H=3$~T from Table~\ref{tab:CW}.  The theory is seen to be in semiquantitative agreement with the experimental data.

\subsection{\label{Sec:EuCu2Sb2HC}Heat Capacity}

\begin{figure}
\includegraphics[width=3in]{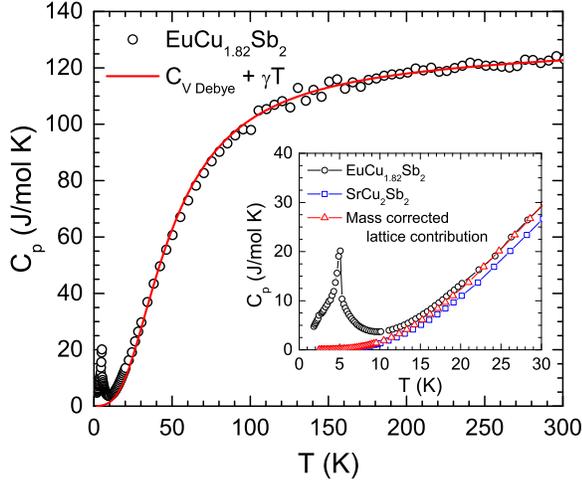}
\caption{(Color online) Heat capacity $C_{\rm p}$ of an \cusbB\ single crystal as a function of temperature $T$ in the temperature range 1.8--300~K measured in zero magnetic field. The solid curve is the fitted sum of the contributions from the Debye lattice heat capacity $C_{\rm V\,Debye}(T)$ and electronic heat capacity $\gamma T$ according to Eq.~(\ref{eq:Debye_HC-fit}). Inset: Expanded view of the low-$T$ $C_{\rm p}(T)$ of \cusbB, of the nonmagnetic reference compound ${\rm SrCu_2Sb_2}$,\cite{Anand2012} and of estimated lattice contribution after correcting for the difference in formula masses of ${\rm EuCu_2Sb_2}$ and \cusbB.}
\label{fig:HC_EuCu2Sb2}
\end{figure}

The heat capacity at constant pressure $C_{\rm p}$ for an \cusbB\ crystal as a function of $T$ is shown in Fig.~\ref{fig:HC_EuCu2Sb2}. A pronounced $\lambda$-type anomaly at 5.1~K (inset of Fig.~\ref{fig:HC_EuCu2Sb2}) is observed that confirms the intrinsic nature of AFM ordering in this compound. Due to the strong influence of the AFM order below $T_{\rm N}$ and short-range AFM order above $T_{\rm N}$ it is difficult to estimate the electronic contribution to $C_{\rm p}(T)$. The strong magnetic contribution at low temperatures prevented us from carrying out a conventional $C/T$ versus $T^2$ fit to the $C/T$ versus $T^2$ data to obtain the Sommerfeld coefficient $\gamma$ and the Debye $T^3$ coefficient $\beta$.

From Fig.~\ref{fig:HC_EuCu2Sb2}, at $T=300$~K the $C_{\rm p}$ attains a value of $\approx123$~J/mol\,K which is close to the expected classical Dulong-Petit value $C_{\rm V} = 3nR = 14.46R = 120.2$~J/mol\,K at constant volume,\cite{Kittel2005, Gopal1966} where $n = 4.82$ is the number of atoms per formula unit (f.u.) and $R$ is the molar gas constant. We analyzed the $C_{\rm p}(T)$ data in the PM state within the framework of the Debye lattice heat capacity model. We fitted the $C_{\rm p}(T)$ data by
\bse
\label{Eqs:CVFit}
\begin{equation}
C_{\rm p}(T) = \gamma T + n C_{\rm{V\,Debye}}(T),
\label{eq:Debye_HC-fit}
\end{equation}
where $C_{\rm{V\,Debye}}(T)$ represents the Debye lattice heat capacity. The Debye model describes the lattice heat capacity due to acoustic phonons at constant volume V which is given per mole of atoms by \cite{Gopal1966}
\begin{equation}
C_{\rm{V\,Debye}}(T) = 9 R \left( \frac{T}{\Theta_{\rm{D}}} \right)^3 {\int_0^{\Theta_{\rm{D}}/T} \frac{x^4 e^x}{(e^x-1)^2}\,dx}.
\label{eq:Debye_HC}
\end{equation}
\ese
The solid curve in Fig.~\ref{fig:HC_EuCu2Sb2} represents the fit of the $C_{\rm p}(T)$ data for ${\rm 15~K\leq T\leq 300~K}$ by Eqs.~(\ref{Eqs:CVFit}) which is obtained using the analytic Pad\'e approximant fitting function for $C_{\rm V\,Debye}(T)$.\cite{Ryan2012} The fit gave the Sommerfeld coefficient $\gamma = 17(2)$~mJ/mol\,K$^2$ and the Debye temperature $\Theta_{\rm D} = 196(2)$~K\@.

\begin{figure}
\includegraphics[width=3in]{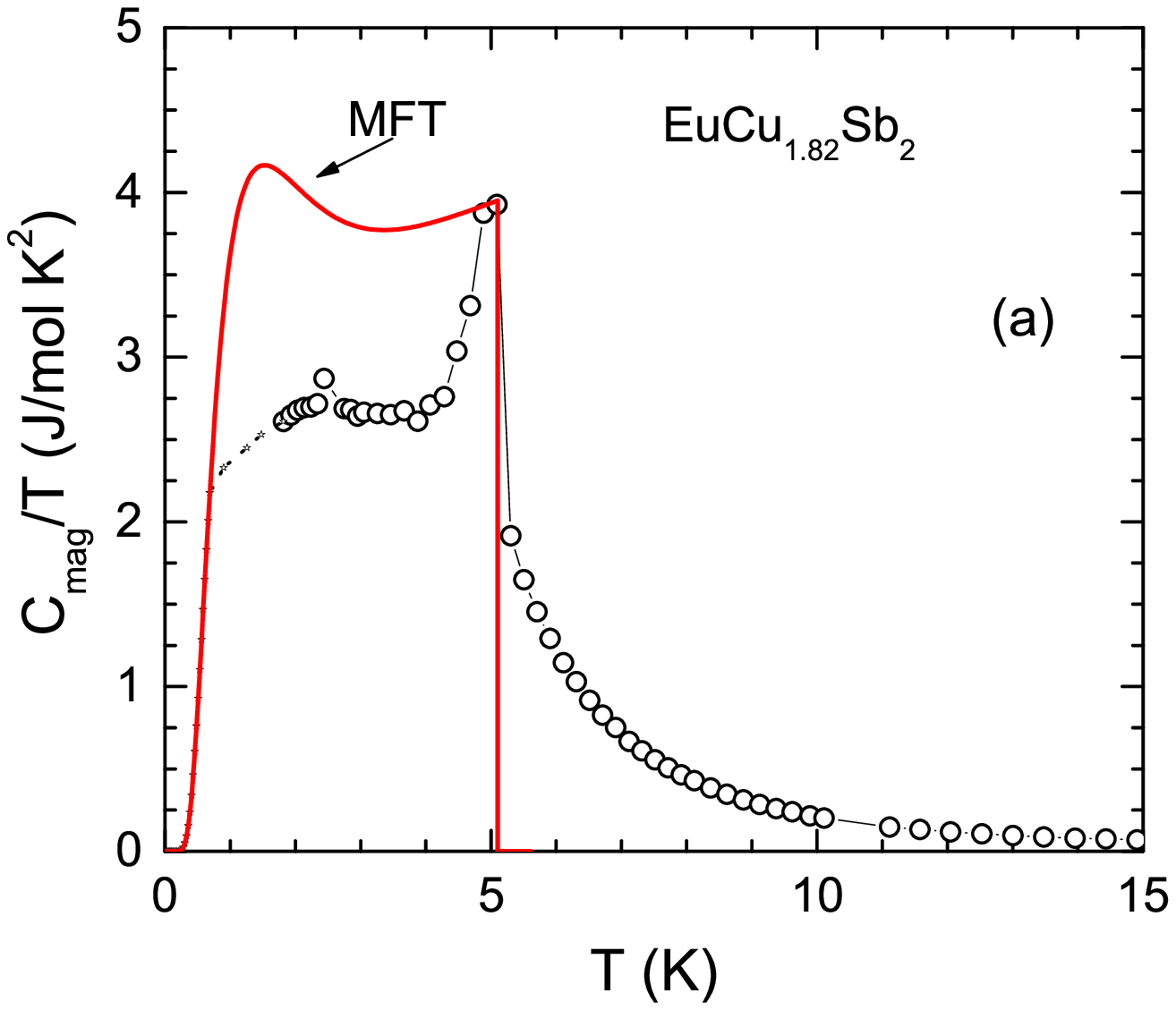}\vspace{0.1in}
\includegraphics[width=3in]{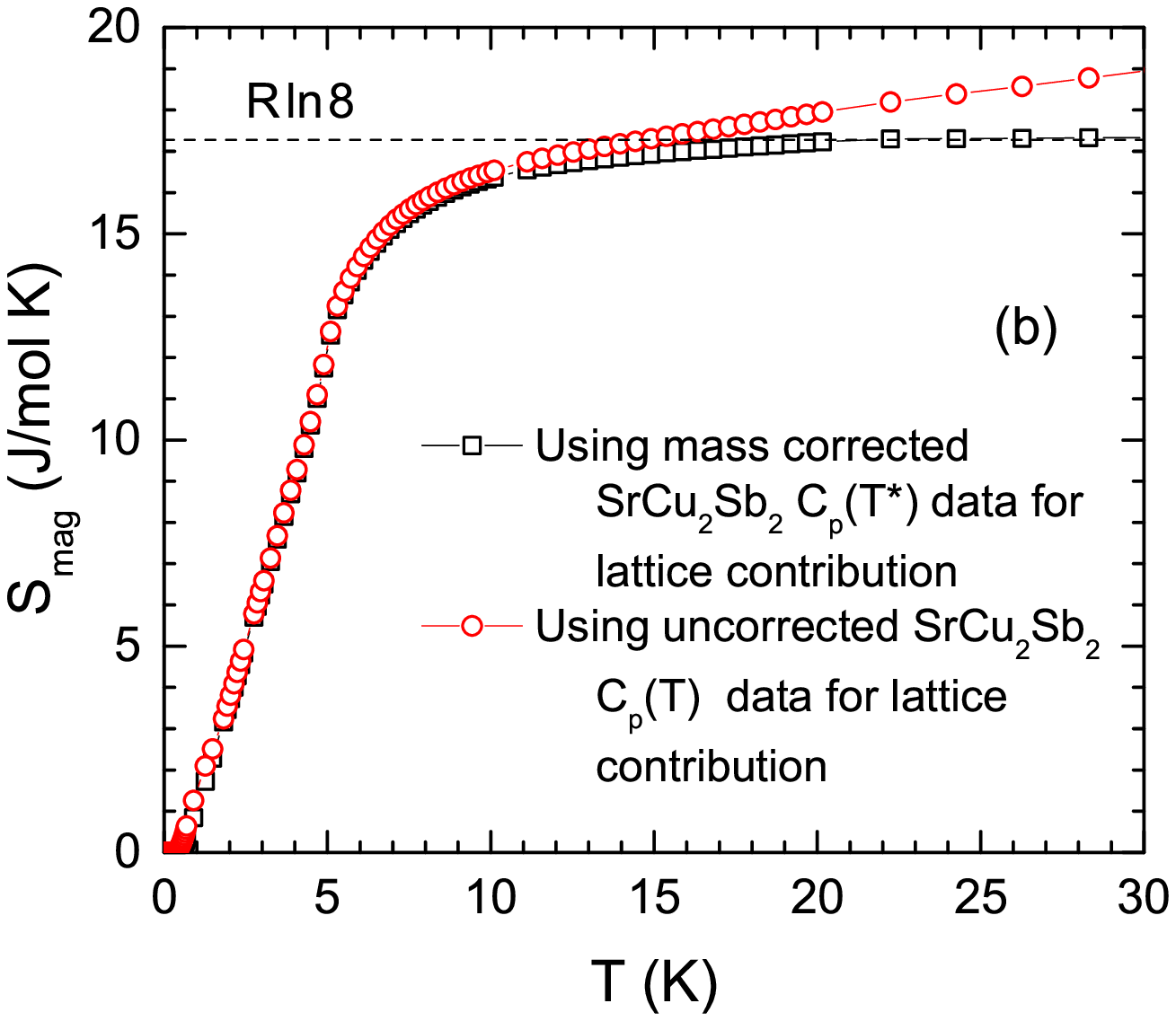}
\caption{(Color online) (a) Magnetic contribution $C_{\rm mag}$ to the heat capacity of \cusbB\ plotted as $C_{\rm mag}(T)/T$ versus $T$\@. The solid curve represents the MFT prediction for $S = 7/2$ and $T_{\rm N}= 5.1$~K\@. (b) Magnetic contribution $S_{\rm mag}$ to the entropy versus $T$\@.  Shown are the data before (red empty circles) and after (black empty squares) formula-mass correction for the lattice heat capacity.}
\label{fig:HC_EuCu2Sb2_mag}
\end{figure}

Now we estimate the magnetic contribution $C_{\rm mag}(T)$ to the heat capacity of \cusbB. For this we used the $C_{\rm p}(T)$ data of ${\rm SrCu_2Sb_2}$ to estimate the lattice contribution to the heat capacity of \cusbB. ${\rm SrCu_2Sb_2}$ also forms in the same ${\rm CaBe_2Ge_2}$-type primitive tetragonal structure as \cusbB. \cite{Anand2012} However, they have different formula masses, therefore the $C_{\rm p}(T)$ data of ${\rm SrCu_2Sb_2}$ need to be corrected for the mass difference in order to obtain an accurate estimate of the lattice heat capacity of \cusbB. As seen from Eq.~(\ref{eq:Debye_HC}), the lattice heat capacity is a function of $T/\Theta_{\rm D}$ and $\Theta_{\rm D}$ depends on the formula mass $M$ according to $\Theta_{\rm D} \sim 1/M^{1/2}$, or equivalently, $T/\Theta_{\rm D}\sim M^{1/2}$. Therefore the measured $T$ of ${\rm SrCu_2Sb_2}$ was scaled as
\begin{equation}
T^* = \frac{T}{(M_{\rm EuCu_2Sb_2}/M_{\rm SrCu_2Sb_2})^{1/2}}.
\label{eq:HC-DebyeT-scaling}
\end{equation}
The mass-corrected lattice contribution thus obtained for \cusbB\ is shown in the inset of Fig.~\ref{fig:HC_EuCu2Sb2}. The $C_{\rm mag}(T)$ of \cusbB\ obtained by subtracting the lattice contribution from the measured $C_{\rm p}(T)$ data of \cusbB\  is shown in Fig.~\ref{fig:HC_EuCu2Sb2_mag}(a) as $C_{\rm mag}(T)/T$ versus $T$, where a sharp anomaly due to the AFM transition is evident. Further we observe that the $C_{\rm mag}(T)$ is nonzero even above the $T_{\rm N}$ and becomes negligible above about 15~K\@. This nonzero contribution to $C_{\rm mag}$ for $T_{\rm N} \leq T \lesssim 15$~K reflects the presence of dynamic short-range AFM correlations above $T_{\rm N}$.

In an AFM state, spin-wave theory predicts a $T^3$ behavior of $C_{\rm mag}(T)$ for a three-dimensional (3D) AFM with negligible anisotropy gap or a $T^2$ dependence in (quasi-) 2D\@.  However, below $T_{\rm N}$ there is significant deviation from the expected power-law behavior.  Instead, a plateau is observed in $C_{\rm mag}(T)/T$ versus $T$ [see Fig.~\ref{fig:HC_EuCu2Sb2_mag}(a)]. This plateau in $C_{\rm mag}(T)/T$ or a corresponding broad  hump in $C_{\rm mag}(T)$ below $T_{\rm N}$ has been observed in several Eu$^{+2}$ and Gd$^{+3}$ compounds, such as in Gd$_2$Fe$_3$Si$_5$ (Ref.~\onlinecite{Vining1983}), GdCu$_2$Si$_2$ (Ref.~\onlinecite{Bouvier1991}), EuB$_6$ (Ref.~\onlinecite{Sullow1998}), EuRh$_2$Si$_2$ (Ref.~\onlinecite{Hossain2001}), and EuCo$_2$Ge$_2$ (Ref.~\onlinecite{Hossain2003}). Attempts have been made in past to explain the origin of the hump in $C_{\rm p}(T)$ below $T_{\rm N}$.\cite{Vining1983, Blanco1991, Fishman1989, Sullow1998} In order to explain the $T$-linear behavior of ordered state $C_{\rm p}(T)/T$ data of Gd$_2$Fe$_3$Si$_5$ Vining and Shelton \cite{Vining1983} speculated that this kind of behavior may be due to low-dimensional spin-wave type ordering. S\"{u}llow et al. \cite{Sullow1998} interpreted the observation of plateau in $C_{\rm p}(T)/T$ of EuB$_6$ in terms of splitting of ground state multiplet of Eu$^{+2}$ by internal magnetic field. According to Fishman and Liu,\cite{Fishman1989} within the Heisenberg model quantum fluctuations can give rise to a hump in $C_{\rm p}(T)$ of a FM system below $T_{\rm N}$.

A hump in $C_{\rm mag}(T)$ arises naturally in the MFT of the ordered-state heat capacity of system having a ($2S+1$)-fold degenerate ground state for large $S$.\cite{Johnston2011, Blanco1991} The origin of the hump can be understood in terms of the entropy that increases with increasing $S$. In order to accommodate the increased entropy a hump appears in the heat capacity. The temperature $T^*$ at which hump develops decreases with increasing $S$. As shown in Ref.~\onlinecite{Johnston2011}, for $S = 7/2$ there appears a hump in $C_{\rm mag}(T)$ data at $T^* \lesssim T_{\rm N}/3$. The solid curve in Fig.~\ref{fig:HC_EuCu2Sb2_mag}(a) represents the mean-field theoretical $C_{\rm mag}(T)/T$ calculated for $T_{\rm N}= 5.1$~K (Ref.~\onlinecite{Johnston2011}) which reproduces the trend of the data. The reduced experimental value of $C_{\rm mag}(T)/T$ below $T_{\rm N}$ is due to the presence of short-range ordering above $T_{\rm N}$ as discussed next.

The experimental magnetic contribution to the entropy $S_{\rm mag}(T)$ was estimated by integrating the $C_{\rm mag}(T)/T$ versus $T$ data according to
\begin{equation}
S_{\rm mag}(T) = {\int_0^T} \frac {C_{\rm mag}(T)}{T} dT.
\label{eq:Smag}
\end{equation}
The $T$ dependence of $S_{\rm mag}$ for \cusbB\  is shown in Fig.~\ref{fig:HC_EuCu2Sb2_mag}(b) for the temperature range $0~{\rm K} \leq T\leq 30$~K\@. The $S_{\rm mag}(T)$ between 0 and 1.8~K was obtained by extrapolating $C_{\rm mag}(T)/T$ to $T=0$ assuming the MFT behavior as shown by the dotted curve in Fig.~\ref{fig:HC_EuCu2Sb2_mag}(a). It is seen from Fig.~\ref{fig:HC_EuCu2Sb2_mag}(b) that if the $S_{\rm mag}(T)$ is uncorrected for the lattice contribution, the obtained $S_{\rm mag}$ exceeds the expected $S_{\rm mag} = R\ln(2S+1) = R\ln8$ for a divalent Eu ($S=7/2$). This illustrates that the lattice contribution must be corrected for to obtain accurate estimates of the magnetic entropy above a few Kelvins. The corrected $S_{\rm mag}$ attains a value of 16.5~J/mol\,K at 10~K which is 95\,\% of the expected high-$T$ limit $R\ln8$. The magnetic entropy of $R\ln8$ is fully recovered at 20~K\@. Thus, consistent with the $\chi(T)$ data, the  $S_{\rm mag}(T)$ data show that the Eu atoms are in the Eu$^{+2}$ state with $S = 7/2$ in \cusbB.  The missing entropy above the MFT curve in Fig.~\ref{fig:HC_EuCu2Sb2_mag}(a) at $T = T_{\rm N}$ is due to the entropy associated with short-range magnetic ordering above $T_{\rm N}$.

\begin{figure}
\includegraphics[width=0.95\columnwidth]{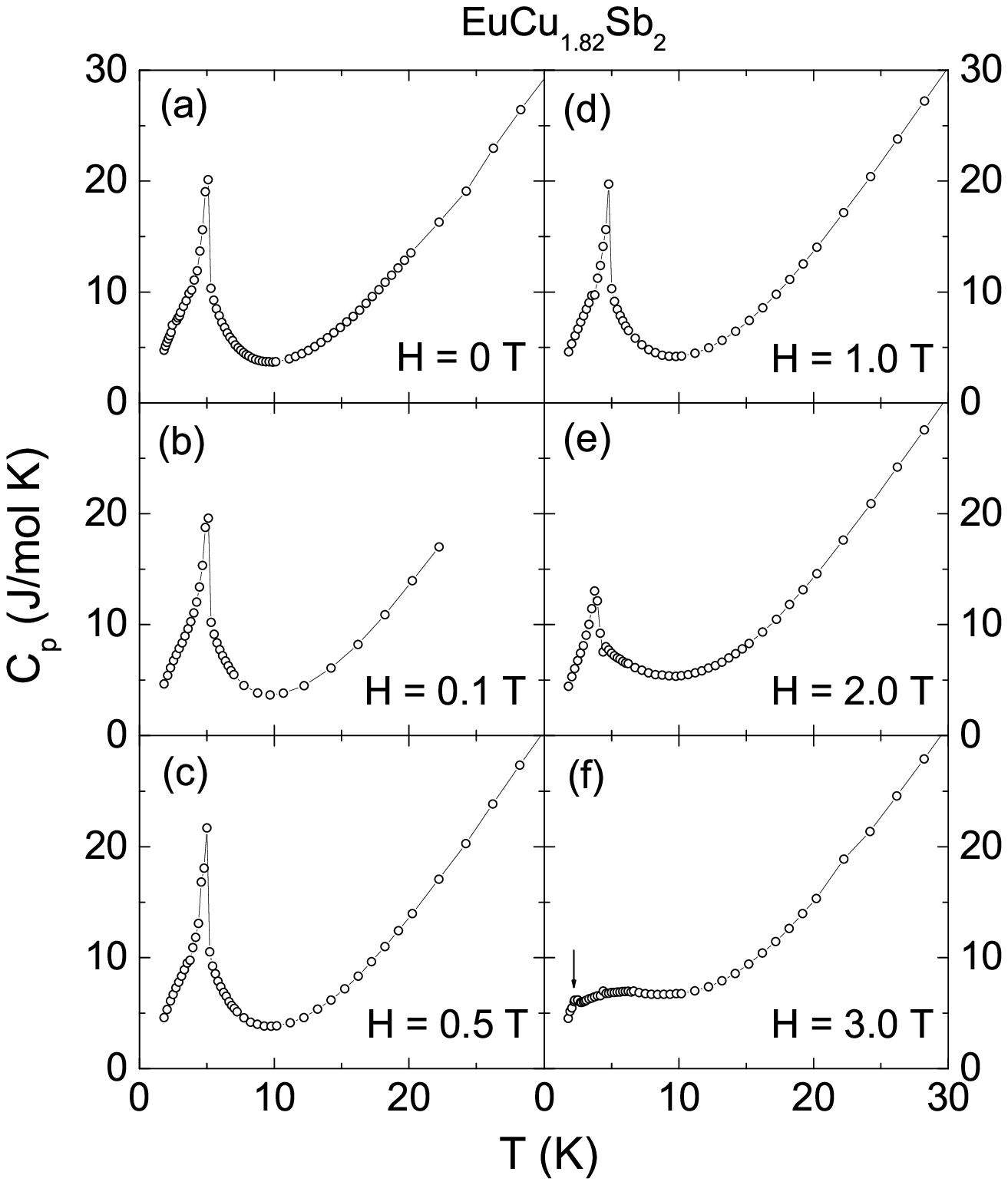}
\caption{(Color online) Heat capacity $C_{\rm p}$ of an \cusbB\ single crystal as a function of temperature $T$ measured in different magnetic fields $H$ applied along the $c$~axis.}
\label{fig:HC_EuCu2Sb2_field}
\end{figure}

\begin{figure}
\includegraphics[width=3in]{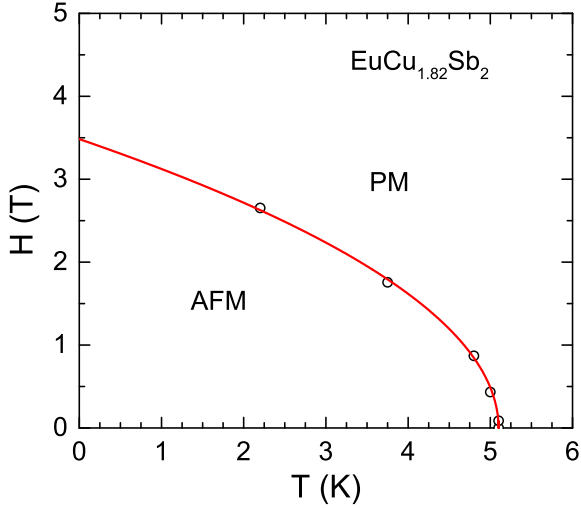}
\caption{(Color online) Magnetic phase diagram in the $H-T$ plane for \cusbB\ as determined from the $C_{\rm p}(H,T)$ data for a single crystal in Fig.~\ref{fig:HC_EuCu2Sb2_field}, where $H$ is the internal field along the $c$~axis.  The solid red curve is a fit of the empirical function $H=H_0\big(1-\frac{T}{T_{\rm N}}\big)^{1/2}$ to the data.}
\label{fig:H-T_EuCu2Sb2}
\end{figure}

We also measured $C_{\rm mag}(T)$ of \cusbB\ in various magnetic fields $H$ applied along the $c$~axis as shown in Fig.~\ref{fig:HC_EuCu2Sb2_field}. Consistent with the $\chi^{-1}(T)$ measurements versus~$H$ in the upper insets of Fig.~\ref{fig:MT_EuCu2Sb2}, the $T_{\rm N}$ decreases with increasing $H$\@. For example, from Fig.~\ref{fig:HC_EuCu2Sb2_field}(e) at $H=2.0$~T, the $T_{\rm N}$ occurs at 3.8~K which as expected for an AFM transition is lower than the zero-field $T_{\rm N}= 5.1$~K\@. It is also seen that the heat capacity jump at $T_{\rm N}$ decreases with increase in $H$ indicating a second order phase transition.  The $H-T$ phase diagram determined from the $H$ dependence of $T_{\rm N}$ from $C_{\rm p}(T)$ measurements in different $H$ is shown in Fig.~\ref{fig:H-T_EuCu2Sb2}, where the solid red curve is a fit of the function $H=H_0\big(1-\frac{T}{T_{\rm N}}\big)^{1/2}$ to the data. The extrapolated critical field at $T=0$ is $\approx3.5$~T\@.

\subsection{\label{Sec:EuCu2Sb2Rho} Electrical Resistivity}

\begin{figure}
\includegraphics[width=3in]{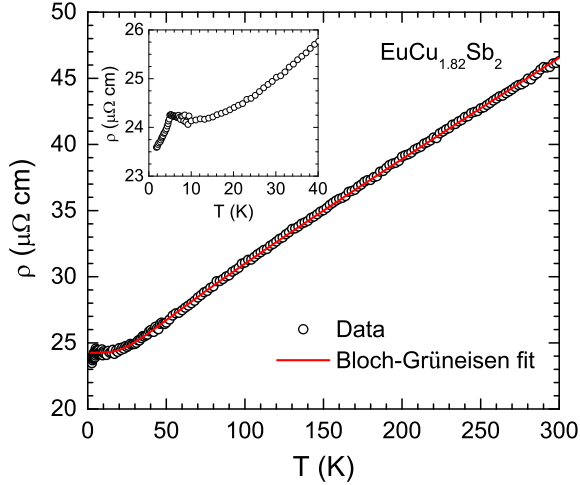}
\caption{(Color online) In-plane electrical resistivity $\rho$ of an \cusbB\ single crystal as a function of temperature $T$ measured in zero magnetic field. The solid curve represents the fit by the Bloch-Gr\"{u}neisen model in Eqs.~(\ref{Eqs:BGModel}) for 6~K~$\leq T \leq$~300~K\@. Inset:~Expanded plot of the low-$T$ $\rho$ data. }
\label{fig:rho_EuCu2Sb2}
\end{figure}

The in-plane $\rho$  of \cusbB\ as a function of $T$ measured in $H=0$ are shown in Fig.~\ref{fig:rho_EuCu2Sb2}. The magnitude and $T$~dependence of $\rho$ indicate metallic behavior. The $\rho$ decreases almost linearly with decreasing $T$ down to 20~K below which it tends to be constant, and eventually meets an AFM transition at $T = 5.1$~K as can be seen from the inset of Fig.~\ref{fig:rho_EuCu2Sb2}, leading to a rapid decrease in $\rho$ below $T_{\rm N}$ due to loss of spin-disorder scattering. For our crystal we find the residual resistivity at $T= 1.8$~K to be $\rho_0 = 23.6~\mu \Omega\,{\rm cm}$ with a residual resistivity ratio ${\rm RRR} \equiv \rho(300\,{\rm K})/\rho(1.8\,{\rm K})\approx 2$.

We analyzed the normal-state $\rho(T)$ data using the Bloch-Gr\"{u}neisen model for the resistivity arising from scattering of electrons from acoustic phonons, \cite{Blatt1968}
\bse
\label{Eqs:BGModel}
\begin{equation}
\rho_{\rm {BG}}(T)= 4 \mathcal{R}(\Theta _{\rm R}) \left( \frac{T}{\Theta _{\rm{R}}}\right)^5 \int_0^{\Theta_{\rm{R}}/T}{\frac{x^5}{(e^x-1)(1-e^{-x})}dx},
 \label{eq:BG}
\end{equation}
\noindent
where $\Theta_{\rm{R}}$ is the Debye temperature determined from resistivity data and $\mathcal{R}(\Theta_{\rm R})$ is a material-dependent constant.  The $\rho(T)$ data for \cusbB\ were fitted by
\begin{equation}
\rho(T) = (\rho_0 + \rho_{\rm sd}) + \rho(\Theta_{\rm R}) f(T/\Theta_{\rm R}),
\label{eq:BG_fit}
\end{equation}
\noindent where $(\rho_0 + \rho_{\rm sd})$ is the sum of the residual resistivity $\rho_0$ due to static defects in the crystal lattice and the spin-disorder resistivity $\rho_{\rm sd}$ due to the presence of disordered magnetic moments. The function $f(y)$ of $y=T/\Theta_{\rm R}$ is defined by \cite{Anand2012, Ryan2012}
\begin{equation}
\begin{split}
f(y) & = \frac{\rho_{\rm BG}(T)}{\rho_{\rm BG}(T=\Theta_{\rm R})} \\
& = 4.226\,259 \,y^5 \int_0^{1/y}\frac{x^5}{(e^x - 1)(1-e^{-x})}\,dx
\label{eq:BG_fn}
\end{split}
\end{equation}
where at $T= \Theta _{\rm{R}}$ the electrical resistivity $\rho(\Theta_{\rm{R}})$ is
\begin{equation}
\rho_{\rm BG}(T=\Theta_{\rm R})=0.9\,464\,635\,{\cal R}(\Theta _{\rm R}).
\label{eq:BG_R}
\end{equation}
\ese

Thus to fit the $\rho(T)$ data requires three independent fitting parameters $(\rho_0 + \rho_{\rm sd})$, $\rho(\Theta_{\rm R})$ and $\Theta_{\rm R}$ using Eqs.~(\ref{eq:BG_fit}) and (\ref{eq:BG_fn}). The solid curve in Fig.~\ref{fig:rho_EuCu2Sb2} shows the fit of the $\rho(T)$ data by Eqs.~(\ref{eq:BG_fit}) and (\ref{eq:BG_fn}) for 6~K~$\leq T \leq$~300~K using the analytic Pad\'e approximant fitting function for $f(y)$.\cite{Ryan2012} The fitted parameters are $\rho_0 + \rho_{\rm sd} = 24.26(2)~\mu \Omega$\,cm, $\rho(\Theta_{\rm{R}}) = 10.2(2)~\mu \Omega$\,cm, and $\Theta_{\rm{R}} = 143(2)$~K. The $\mathcal{R}(\Theta _{\rm R})$ can be calculated from the value of $\rho(\Theta_{\rm{R}})$ using Eq.~(\ref{eq:BG_R}), yielding $\mathcal{R}(\Theta _{\rm R}) = 10.8~\mu \Omega$\,cm. Then $\rho_{\rm sd}$ is calculated from the value of $\rho_0 + \rho_{\rm sd}$ using the above 1.8~K value $\rho_0 = 23.6~\mu \Omega$\,cm which gives $\rho_{\rm sd} \approx 0.7~\mu \Omega$\,cm.

\section{\label{EuCu2As2} Physical Properties of E\lowercase{u}C\lowercase{u}$_2$A\lowercase{s}$_2$ Crystals}

\subsection{\label{Sec:EuCu2As2ChiMH} Magnetization and Magnetic Susceptibility}

\begin{figure}
\includegraphics[width=3in]{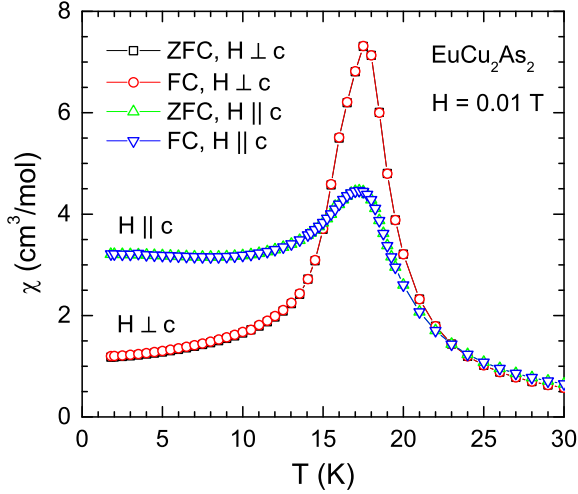}
\caption{(Color online) Zero-field-cooled (ZFC) and field-cooled (FC) magnetic susceptibility $\chi$ of an ${\rm EuCu_2As_2}$ single crystal as a function of temperature $T$ in the temperature range 1.8--30~K measured in a magnetic field $H= 0.01$~T applied along the $c$~axis ($\chi_c, H \parallel c$) and in the $ab$~plane ($\chi_{ab}, H \perp  c$).}
\label{fig:MT_EuCu2As2_low-H}
\end{figure}

\begin{figure}
\includegraphics[width=3in]{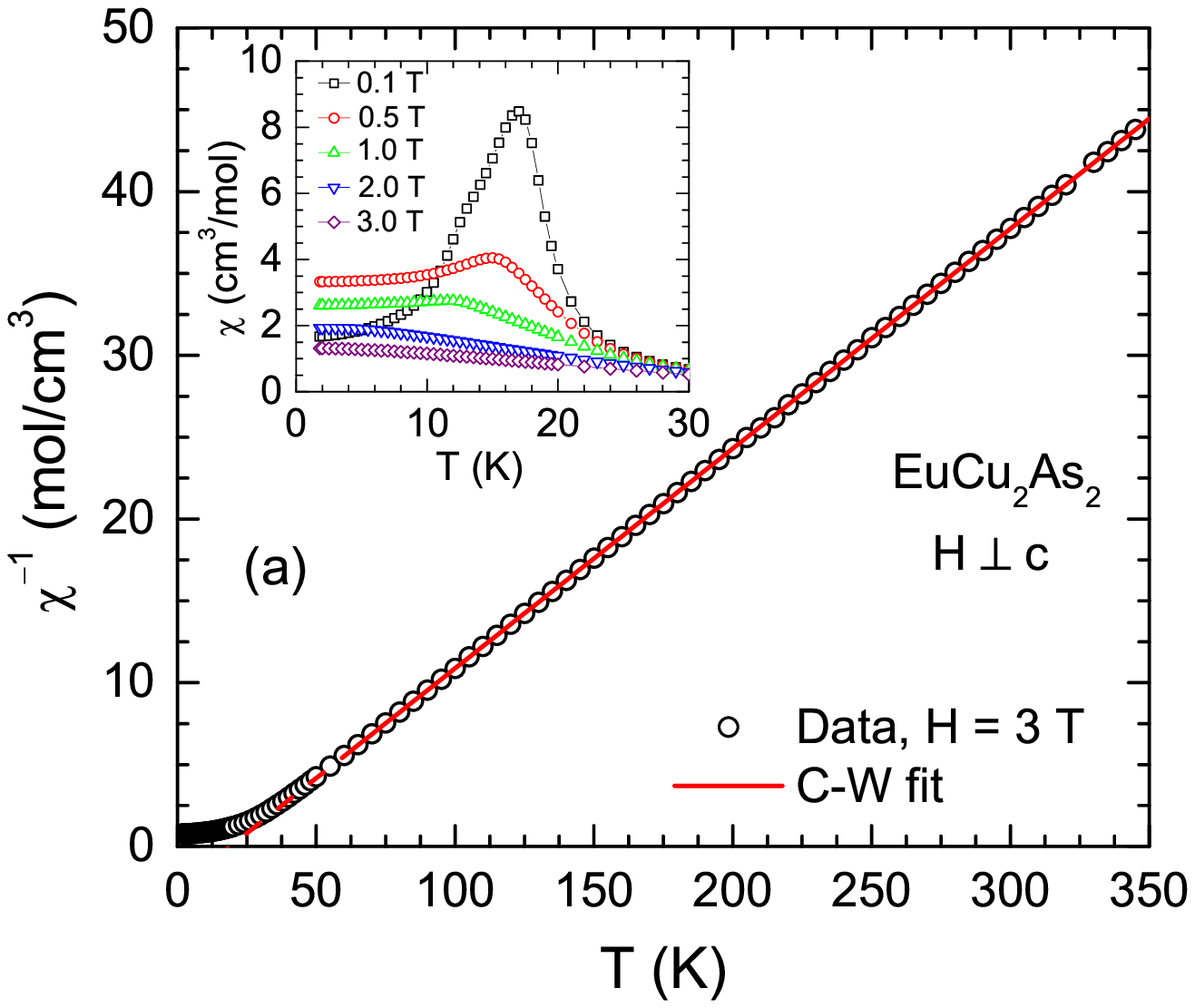}\vspace{0.1in}
\includegraphics[width=3in]{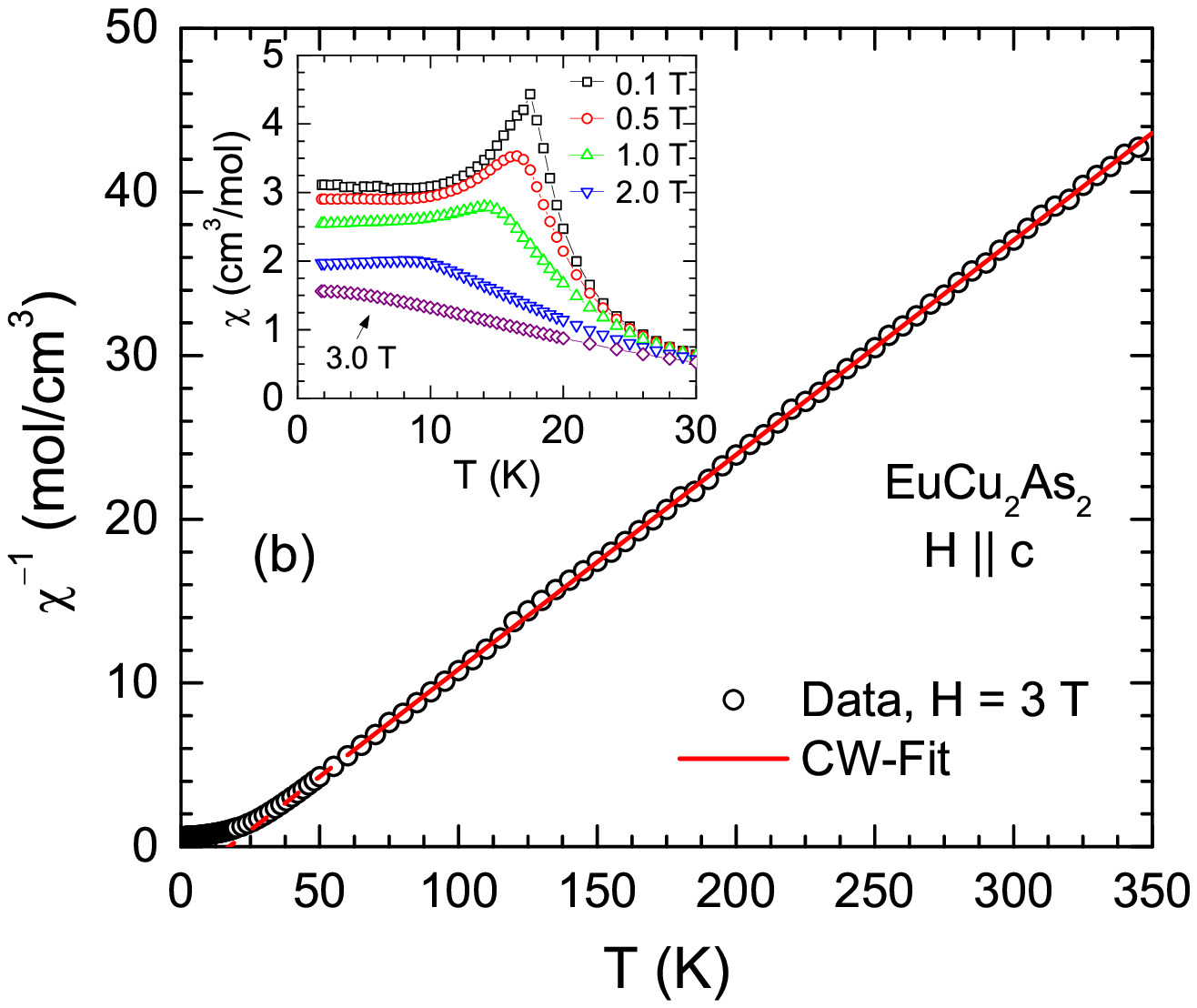}
\caption{(Color online) Zero-field-cooled inverse magnetic susceptibility $\chi^{-1}$ of an ${\rm EuCu_2As_2}$ single crystal (crystal~1) versus temperature $T$ in the $T$ range 1.8--350~K measured in a magnetic field $H=3$~T applied (a) in the $ab$~plane ($\chi_{ab}, H \perp  c$) and, (b) along the $c$~axis ($\chi_c, H \parallel c$). The insets in (a) and (b) show low-$T$ $\chi$ data at different $H$ values.}
\label{fig:MT_EuCu2As2}
\end{figure}

The ZFC and FC $\chi(T)$ data for ${\rm EuCu_2As_2}$ single crystal measured at different $H$ applied along the $c$~axis ($\chi_c,\ H \parallel c$) and in the $ab$~plane ($\chi_{ab},\  H \perp c$) are shown in Figs.~\ref{fig:MT_EuCu2As2_low-H} and \ref{fig:MT_EuCu2As2}. As can be seen from Fig.~\ref{fig:MT_EuCu2As2_low-H}, at low applied $H = 0.01$~T, the $\chi(T)$ data exhibit well pronounced AFM ordering features at $T_{\rm N} = 17.5$~K for both $H \parallel c$ and $H \perp c$ without any hysteresis in the ZFC and FC $\chi(T)$ data. Like \cusbB, the $T_{\rm N}$ of ${\rm EuCu_2As_2}$ decreases with increasing $H$ as shown in the insets of Fig.~\ref{fig:MT_EuCu2As2}.  Our $T_{\rm N}=17.5$~K is significantly larger than the value of 15~K previously reported by Sengupta et al.\cite{Sengupta2005} At $H=0.1$~T we observe a weak change in slope of $\chi(T)$ near 12~K [see inset of Fig.~\ref{fig:MT_EuCu2As2}(a)] for $H \perp c$, the origin of which is not clear.

The $\chi_{ab}(T)$ of ${\rm EuCu_2As_2}$ in Fig.~\ref{fig:MT_EuCu2As2_low-H} exhibits an anomalous strong increase for $T_{\rm N} < T < 20$~K that may be due to the buildup of FM correlations beyond MFT\@. This behavior is not predicted by MFT and is in contrast to the data above $T_{\rm N}$ for \cusbB\ in Fig.~\ref{fig:MT_EuCu2Sb2_low-H}.  It is seen in Fig.~\ref{fig:MT_EuCu2As2_low-H} that $\chi_{ab}/\chi_c \approx 1.7$ at $T\approx T_{\rm N}$. However, at lower $T < 13$~K, $\chi_{ab} < \chi_c$ which indicates that the easy axis or easy plane is perpendicular to the $c$~axis. This easy-plane behavior in ${\rm EuCu_2As_2}$ is the same as observed above in \cusbB.

The $\chi(T)$ data in the PM state above $\sim 50$~K are well represented by the Curie-Weiss law~(\ref{eq:C-W}). Linear fits of $\chi^{-1}(T)$ measured at $H=3$~T in the temperature range 60~K~$\leq T \leq$~350~K are shown by straight red lines in Fig.~\ref{fig:MT_EuCu2As2}. The fitted parameters are $C = 7.44(1)$~cm$^3$\,K/mol and $\theta_{\rm p}^{ab}= +19.0(2)$~K for $\chi_{ab}$ and $C = 7.64(2)$~cm$^3$\,K/mol and $\theta_{\rm p}^c = +17.2(5)$~K for $\chi_c$. The effective moments obtained from the values of $C$ are $\mu_{\rm eff} = 7.72(1)\, \mu_{\rm B}$ from $\chi_{ab}$ and $\mu_{\rm eff} = 7.82(2)\, \mu_{\rm B}$ from $\chi_c$. Again these values of $\mu_{\rm eff}$ are similar to the theoretical value of $7.94\, \mu_{\rm B}$ for a free Eu$^{+2}$ ion with $S=7/2$ and $g=2$, indicating the presence of divalent Eu ions in ${\rm EuCu_2As_2}$.  According to Eq.~(\ref{eq:thetap}), the strongly positive values of $\theta_{\rm p}\sim T_{\rm N}$ indicate the dominance of (negative) FM interactions~$J_{ij}$ over (positive) AFM interactions between the Eu spins in \cuas.

\begin{figure}
\includegraphics[width=3in]{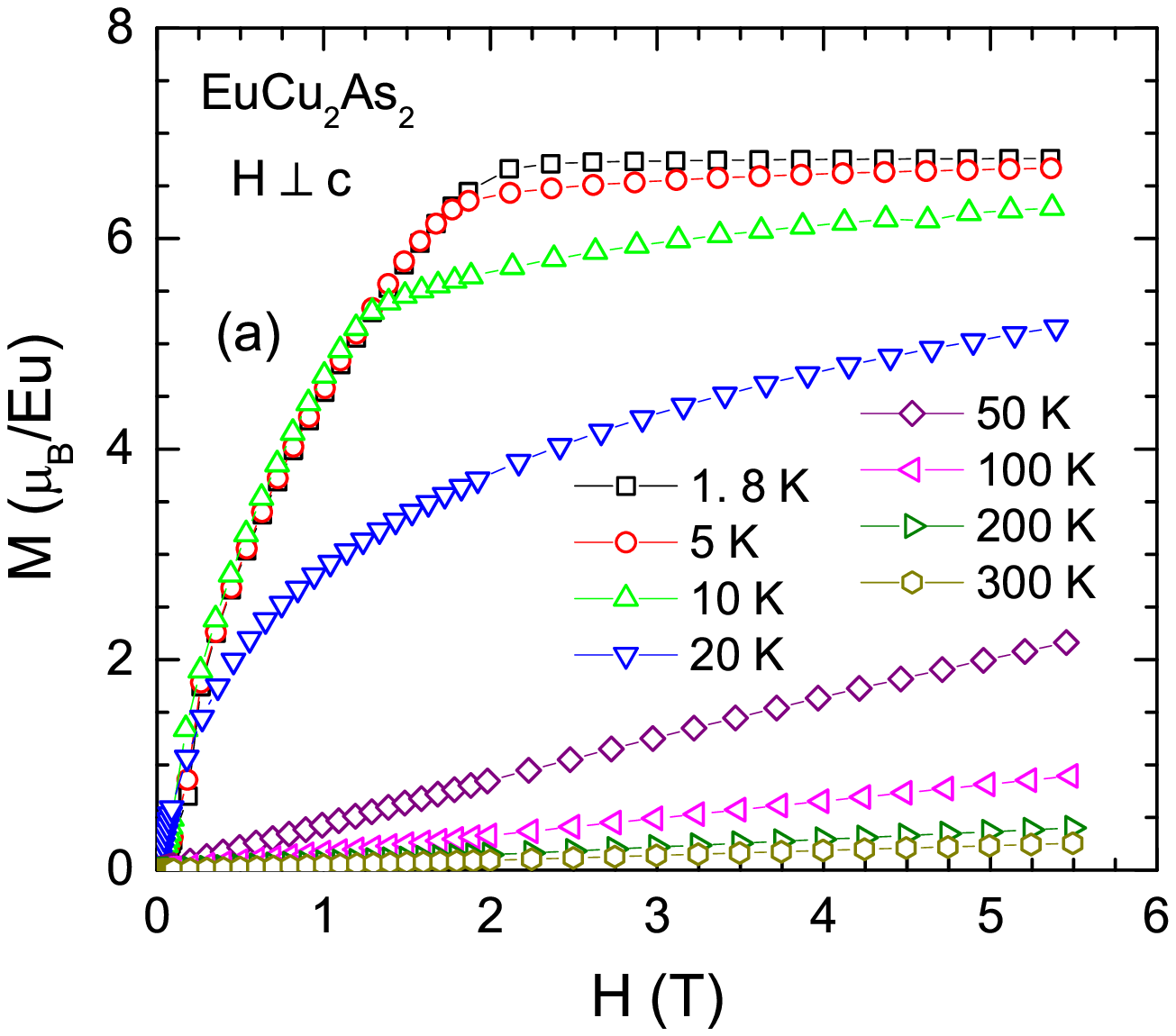}\vspace{0.1in}
\includegraphics[width=3in]{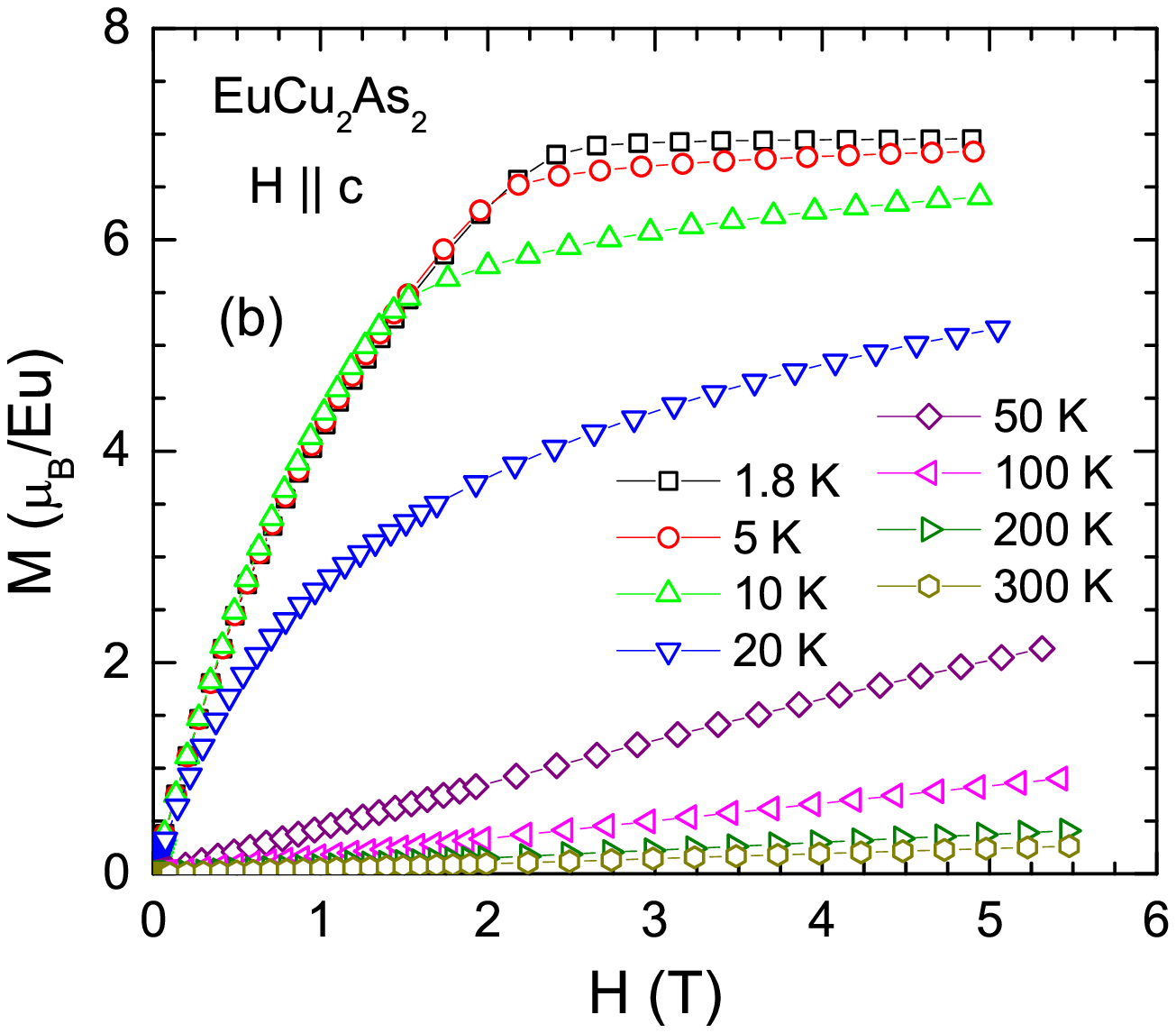}
\caption{(Color online) Isothermal magnetization $M$ of an ${\rm EuCu_2As_2}$ single crystal as a function of internal magnetic field $H$ measured at the indicated temperatures for $H$ (a) in the $ab$~plane ($M_{ab}, H \perp  c$) and, (b) along the $c$~axis ($M_c, H~\parallel~c$).}
\label{fig:MH_EuCu2As2}
\end{figure}

\begin{figure}
\includegraphics[width=3in]{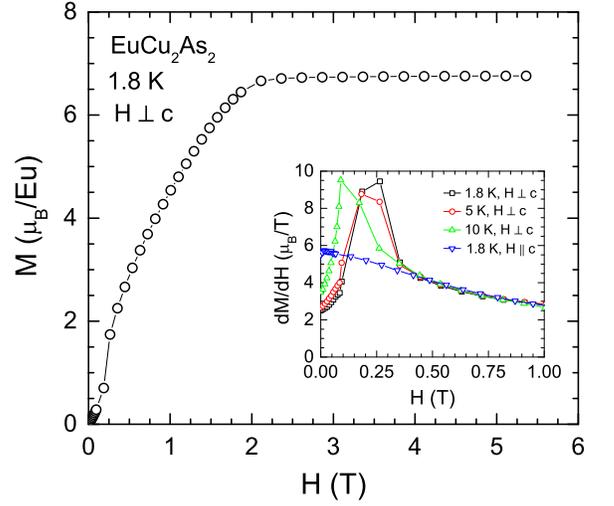}
\caption{(Color online) Isothermal magnetization $M$ of an ${\rm EuCu_2As_2}$ single crystal as a function of internal magnetic field $H$ measured at 1.8~K for $H \perp c$. Inset: The derivative $dM/dH$ vs. $H$ plot at 1.8, 5 and 10~K for $H \perp  c$ and $H \parallel c$.}
\label{fig:MH_EuCu2As2_2K}
\end{figure}

The isothermal $M(H)$ data of the ${\rm EuCu_2As_2}$ crystal measured at different temperatures between 1.8~K and 300~K for $H$ applied along the $c$~axis ($M_c, H \parallel c$) and in the $ab$~plane ($M_{ab}, H \perp  c$) are shown in Fig.~\ref{fig:MH_EuCu2As2}. As $T$ decreases below 20~K, strong negative curvature occurs in $M(H)$ for both $H \parallel c$ and $H \perp c$, again indicating the presence of dominant FM interactions and correlations.  It is striking, however, that even at the lowest $T = 1.8$~K this strong negative curvature persists, in contrast to the prediction of MFT in Fig.~\ref{fig:MH_EuCu2Sb2_MFT} for the perpendicular magnetization for an AFM which should be linear in field for $T < T_{\rm N}$ up to the critical field which from Fig.~\ref{fig:MH_EuCu2As2} is of order 2~T\@.  This strong divergence from the prediction of MFT for a planar AFM suggests that the AFM magnetic structure is both noncollinear and noncoplanar.  The noncoplanar component likely consists of canting of the Eu moments out of the $ab$ plane in an alternating way such that the overall magnetic structure is AFM\@.

At $H=5.5$~T, the saturation moments are $\mu_{\rm sat}^{ab} = 6.76\,\mu_{\rm B}$/f.u.\ and $\mu_{\rm sat}^{c} = 6.95\,\mu_{\rm B}$/f.u., close to the expected value of $\mu_{\rm sat} = gS\mu_{\rm B}$ with $g=2$ and $S=7/2$. The $M(H)$ isotherm at 1.8~K is found to exhibit no magnetic field hysteresis. The $M(H)$ isotherms at 5~K are almost the same as those at 1.8~K\@. With increasing $T$, $\mu_{\rm sat}$ monotonically decreases as one approaches $T_{\rm N}$ as expected.

The $M(H)$ curve at 1.8~K for $H \perp c$ is shown separately in Fig.~\ref{fig:MH_EuCu2As2_2K} to illustrate a spin flop transition at $H\approx0.2$~T which is more clearly seen in the derivative plots ($dM/dH\ {\rm vs.}\ H$) shown in the inset of Fig.~\ref{fig:MH_EuCu2As2_2K}. The slope changes in the $M(H)$ curves for $H \perp c$ are clearly illustrated from the $dM/dH\ {\rm vs.}\ H$ plots at 1.8, 5 and 10~K in the inset of Fig.~\ref{fig:MH_EuCu2As2_2K}. In contrast, no such behavior is observed in $dM/dH\ {\rm vs.}\ H$ plot for $H \parallel c$. This behavior is consistent with the above $\chi(T)$ and $M(H)$ data that indicate that the ordered Eu$^{+2}$ moments are aligned in the $ab$ plane apart from the above suggested canting in and out of that plane.

\subsection{\label{Sec:EuCu2As2HC}Heat Capacity}

\begin{figure}
\includegraphics[width=3in]{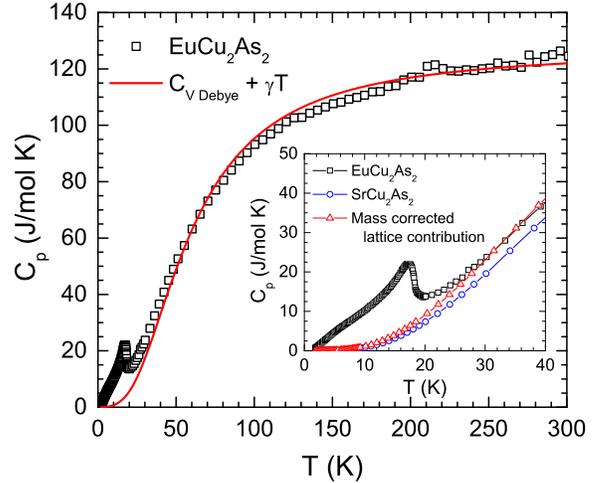}
\caption{(Color online) Heat capacity $C_{\rm p}$ of an ${\rm EuCu_2As_2}$ single crystal as a function of temperature $T$ in the $T$ range 1.8--300~K measured in zero magnetic field. The solid curve is the sum of the contributions from the Debye lattice heat capacity $C_{\rm V\,Debye}(T)$ and electronic heat capacity $\gamma T$ according to Eq.~(\ref{eq:Debye_HC-fit}).  Inset: Expanded view of the low-$T$ $C_{\rm p}(T)$ data and the estimated lattice contribution of ${\rm SrCu_2As_2}$ (Ref.~\onlinecite{Anand2012}) before and after correcting for the difference in formula masses of ${\rm EuCu_2As_2}$ and ${\rm SrCu_2As_2}$.}
\label{fig:HC_EuCu2As2}
\end{figure}

The $C_{\rm p}(T)$ data for ${\rm EuCu_2As_2}$ are shown in Fig.~\ref{fig:HC_EuCu2As2}. The $C_{\rm p}(T)$ exhibits a distinct second-order transition at 17.5~K as shown in the inset of Fig.~\ref{fig:HC_EuCu2As2} due to the AFM ordering. The $C_{\rm p}$ attains a saturation value of $\sim$ 124.5~J/mol\,K at room temperature which is close to the expected classical Dulong-Petit value of $C_{\rm V}$ = 124.7~J/mol\,K\@. The $C_{\rm p}(T)$ data were analyzed using the Debye lattice heat capacity model in the temperature range $30\leq T\leq 300$~K\@. The fit of the $C_{\rm p}(T)$ data by Eqs.~(\ref{Eqs:CVFit}) using the analytic Pad\'e approximant fitting function \cite{Ryan2012} is shown by the solid curve in Fig.~\ref{fig:HC_EuCu2As2} where again we used $\gamma$ and $\Theta_{\rm D}$ as adjustable parameters for the fit with ${\rm 30~K} \leq T\leq 300$~K, yielding $\gamma = 4(3)~{\rm mJ/mol\,K^2}$ and $\Theta_{\rm D} = 241(3)$~K\@.

\begin{figure}
\includegraphics[width=3in]{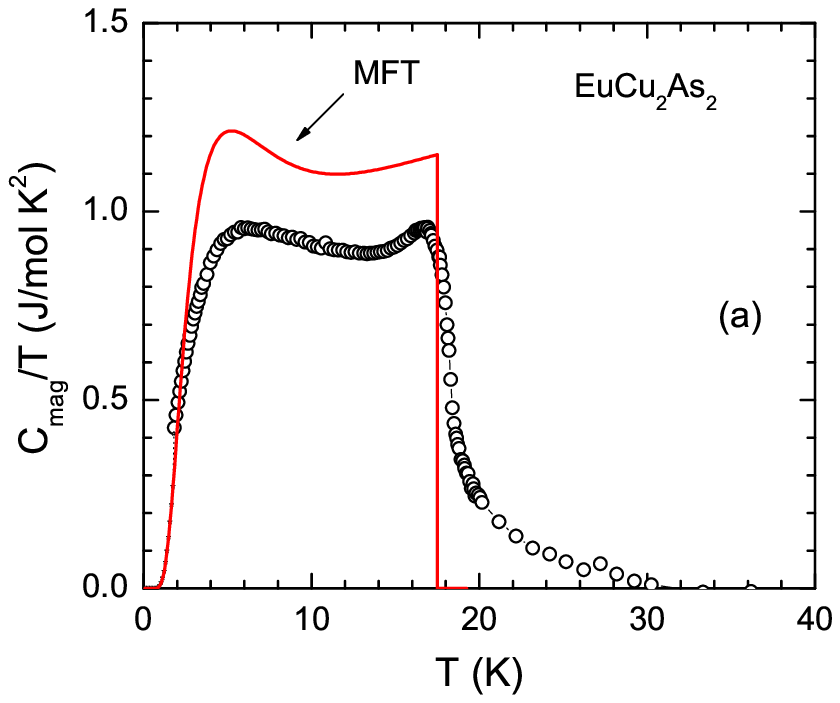}\vspace{0.1in}
\includegraphics[width=3in]{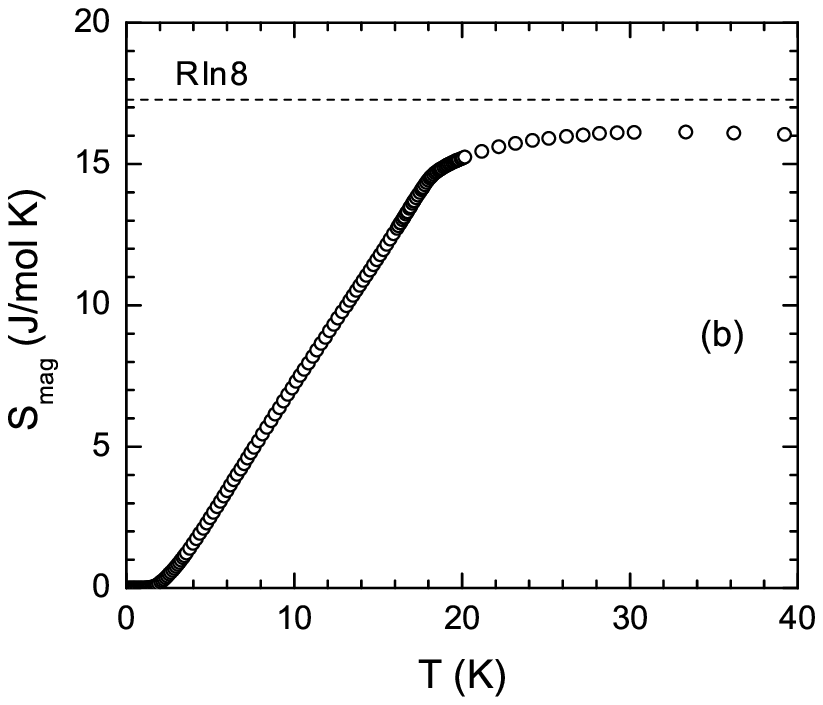}
\caption{(Color online) (a) Magnetic contribution to the heat capacity $C_{\rm mag}$ for ${\rm EuCu_2As_2}$ plotted as $C_{\rm mag}(T)/T$ versus $T$\@. The solid curve represents the MFT prediction of $C_{\rm mag}$ for $S = 7/2$ and $T_{\rm N}= 17.5$~K\@. (b) Magnetic contribution  $S_{\rm mag}(T)$ to the entropy.}
\label{fig:HC_EuCu2As2_mag}
\end{figure}

The $C_{\rm mag}(T)$ and $S_{\rm mag}(T)$ data for ${\rm EuCu_2As_2}$ are shown in Fig.~\ref{fig:HC_EuCu2As2_mag} which were obtained after subtracting the lattice contribution to $C_{\rm p}(T)$ as done above for \cusbB. The mass-normalized lattice contribution for ${\rm EuCu_2As_2}$ is shown in the inset of Fig.~\ref{fig:HC_EuCu2As2}. Here again we observe from Fig.~\ref{fig:HC_EuCu2As2_mag}(a) that $C_{\rm mag}(T)$ is nonzero above the $T_{\rm N}$ up to 30~K due to the presence of magnetic correlations. The appearance of the plateau in $C_{\rm mag}(T<T_{\rm N})/T$ is consistent with MFT. \cite{Johnston2011} The $C_{\rm mag}(T)/T$ calculated for $T_{\rm N}= 17.5$~K using MFT is shown as the solid red curve in Fig.~\ref{fig:HC_EuCu2As2_mag}(a). Here again the presence of magnetic fluctuations well above the $T_{\rm N}$ is manifested as a reduced value of $C_{\rm mag}(T)/T$ below $T_{\rm N}$ compared to the calculated MFT behavior. The $S_{\rm mag}(T)$ data shown in Fig.~\ref{fig:HC_EuCu2As2_mag}(b) were determined by integrating $C_{\rm mag}(T)/T$ versus $T$, and is found to attain a value of 15.6~J/mol\,K ($\approx 0.90\,R\ln8$) at 22~K and 16.1~J/mol\,K ($\approx 0.93\,R\ln8$) at 30~K, thus confirming the divalent state of Eu atoms in ${\rm EuCu_2As_2}$.

\begin{figure}
\includegraphics[width=3in]{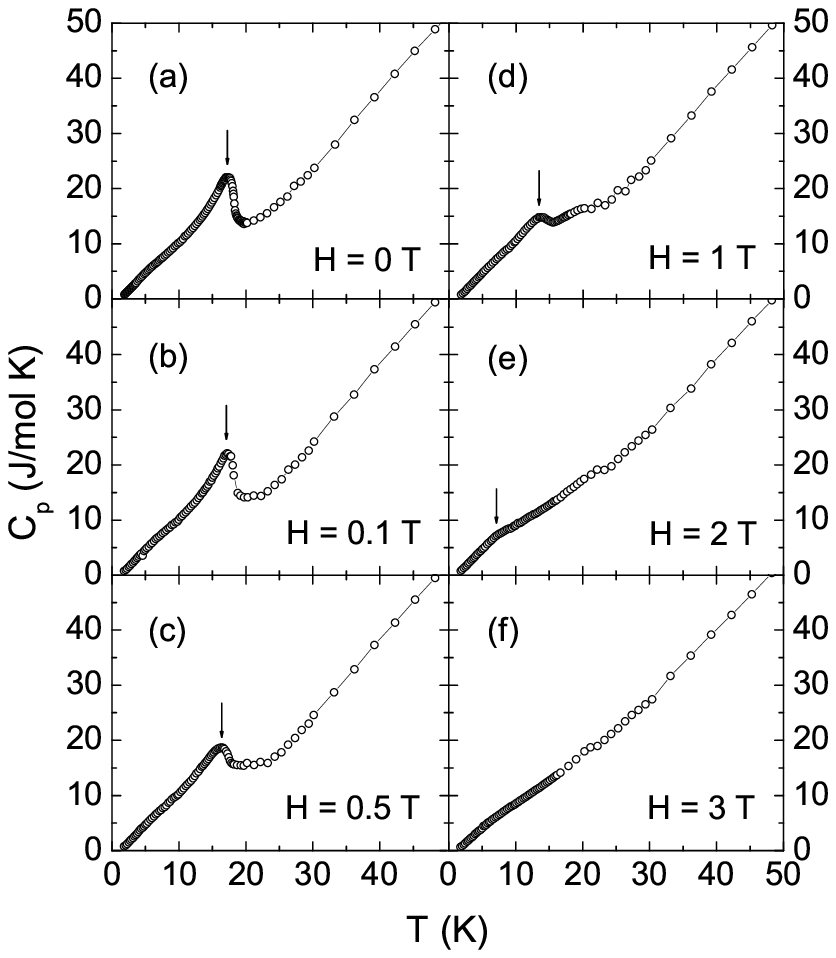}
\caption{Heat capacity $C_{\rm p}$ of an ${\rm EuCu_2As_2}$ single crystal as a function of temperature $T$ measured in different magnetic fields $H$ applied along the $c$~axis. }
\label{fig:HC_EuCu2As2_field}
\end{figure}

\begin{figure}
\includegraphics[width=3in]{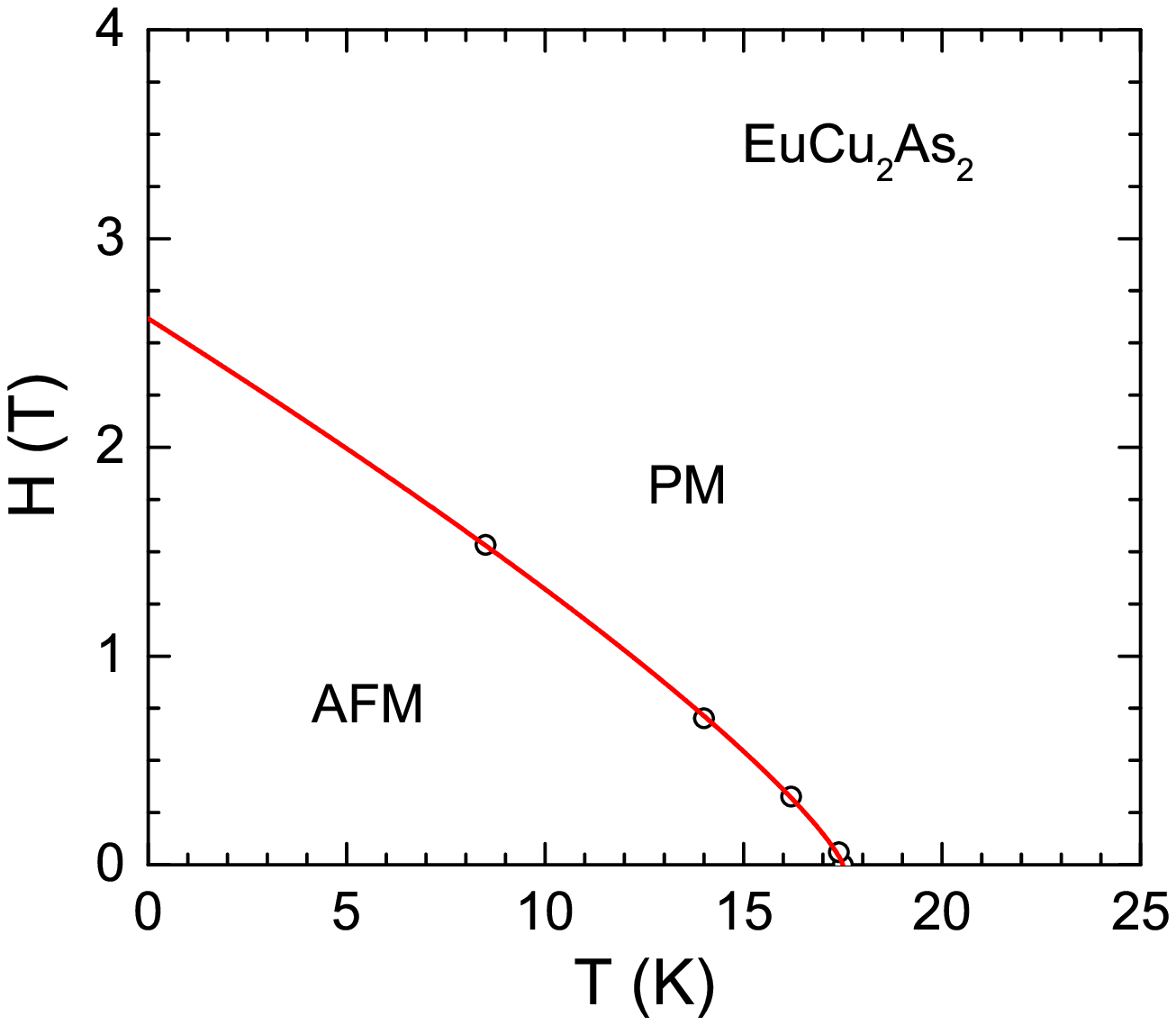}
\caption{Magnetic phase diagram in the $H-T$ plane for \cuas\ as determined from the $C_{\rm p}(H,T)$ data in Fig.~\ref{fig:HC_EuCu2As2_field}, where $H$ is the internal field along the $c$~axis.  The solid red curve is a fit of the empirical function $H=H_0\big(1-\frac{T}{T_{\rm N}}\big)^{0.81}$ to the data. }
\label{fig:H-T_EuCu2As2}
\end{figure}

The $C_{\rm p}(T)$ data of ${\rm EuCu_2As_2}$ measured under the application of different $H$ along the $c$~axis are shown in Fig.~\ref{fig:HC_EuCu2As2_field}. It is seen that $T_{\rm N}$ decreases as $H$  increases.  For example at $H=1.0$~T, the $T_{\rm N} = 13.5$~K compared with $T_{\rm N}= 17.5$~K at $H=0$. At the same time the heat capacity anomaly becomes broader with increasing $H$\@.  We also observe that with increasing $H$ a broad peak appears near 20~K for $H = 1.0$~T [Fig.~\ref{fig:HC_EuCu2As2_field}(d)] which we suggest is due to field-induced FM correlations above $T_{\rm N}(H)$. At $H=3.0$~T no clear transition is observed in the $C_{\rm p}(T)$ data. The $H-T$ magnetic phase diagram determined from the $H$ dependence of $T_{\rm N}$ obtained from the $C_{\rm p}(H,T)$ measurements in Fig.~\ref{fig:HC_EuCu2As2_field} is shown in Fig.~\ref{fig:H-T_EuCu2As2}.  The critical field at which AFM is destroyed at $T=0$~K is estimated from the extrapolated red curve in the figure to be $\approx 2.6$~T\@.  In contrast, for the polycrystalline \cuas\ sample in Ref.~\onlinecite{Sengupta2012} with $T_{\rm N}\approx15$~K the extrapolated critical field is $\approx 1.8$~T\@.  The differences between the $T_{\rm N}$ and critical field values in Refs.~\onlinecite{Sengupta2005} and~\onlinecite{Sengupta2012} and ours are correlated with the fact that their samples were polycrystals whereas ours are single crystals.  The different preparation conditions evidently lead to differences in compositions and/or defect concentrations, resulting in differences in the physical properties.

\subsection{\label{Sec:EuCu2As2Rho} Electrical Resistivity}

\begin{figure}
\includegraphics[width=3in]{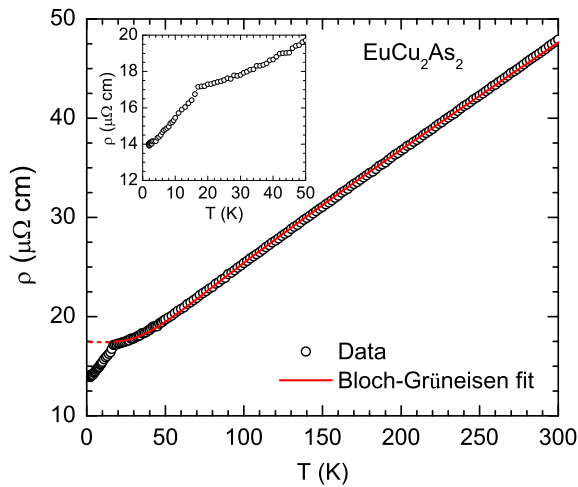}
\caption{(Color online) In-plane electrical resistivity $\rho$ of an ${\rm EuCu_2As_2}$ single crystal versus temperature $T$ in $H=0$. The solid curve is the fit by the Bloch-Gr\"{u}neisen model in Eqs.~(\ref{Eqs:BGModel}) for 18~K~$\leq T \leq$~300~K\@. The dashed curve is an extrapolation of the fit to $T=0$.}
\label{fig:rho_EuCu2As2}
\end{figure}

The in-plane $\rho(T)$ data of ${\rm EuCu_2As_2}$ measured in zero magnetic field are shown in Fig.~\ref{fig:rho_EuCu2As2}. The data exhibit metallic behavior with $\rho_0 = 14.0(1)~\mu \Omega\,{\rm cm}$ at $T= 1.8$~K and ${\rm RRR} \approx 3.5$. As shown in the inset of Fig.~\ref{fig:rho_EuCu2As2}, a sharp transition is seen at $T_{\rm N} = 17.0$~K in the $\rho(T)$ data which decreases at lower $T$ due to a reduction in spin-disorder scattering. The $\rho(T)$ data are well represented by the Bloch-Gr\"{u}neisen model. The fit of the $\rho(T)$ data by Eqs.~(\ref{Eqs:BGModel}) is shown by solid curve in Fig.~\ref{fig:rho_EuCu2As2} in the temperature range 18~K~$\leq T \leq$~300~K using the analytic Pad\'e approximant fitting function for $\rho_{\rm BG}(T)$.\cite{Ryan2012} The fitting parameters are $\rho_0 + \rho_{\rm sd} = 17.44(4)~\mu \Omega$\,cm, $\rho(\Theta_{\rm{R}}) = 21.9(3)~\mu \Omega$\,cm, and $\Theta_{\rm{R}} = 223(3)$~K\@. We obtained $\mathcal{R}(\Theta _{\rm R})$ = 23.1~$\mu \Omega$\,cm using Eq.~(\ref{eq:BG_R}). The $\rho_{\rm sd} \approx 3.4~\mu \Omega$\,cm is obtained from $\rho_0 + \rho_{\rm sd}$ using $\rho_0 = 14.0(1)~\mu \Omega$\,cm at 1.8~K\@.

\section{\label{Conclusion} Summary}

We have investigated the crystallographic, magnetic, thermal and transport properties of ${\rm EuCu_2As_2}$ and \cusbB\ single crystals. The Rietveld refinements of x-ray powder diffraction data for crushed crystals indicate full occupancy of the atomic sites in \cuas, whereas Cu vacancies are found in \cusb\ corresponding to the actual composition \cusbB.  On the other hand, no evidence for Cu vacancies in a polycrystalline sample of \cusb\ were found.\cite{Ryan2015}  This difference evidently reflects differences in the sample type (powder versus single crystal) and preparation conditions. A significant concentration of Cu vacancies was previously found in the similar compound ${\rm CaCu_{1.7}As_2}$.\cite{Anand2012b}

The $T$ dependences of $\rho$ and $C_{\rm p}$ and the magnitude of $\rho$ indicate metallic ground states in both compounds.  The $\rho(T > T_{\rm N})$ and $C_{\rm p}(T > T_{\rm N})$ data were successfully analyzed by the Bloch-Gr\"{u}neisen model and the Debye model of lattice heat capacity, respectively.  The $\chi(T)$, $C_{\rm p}(T)$, and $\rho (T)$ demonstrate the occurrence of long-range AFM order below $T_{\rm N} = 5.1$~K in \cusbB\ and $T_{\rm N} = 17.5$~K in ${\rm EuCu_2As_2}$. The spin $S=7/2$ state of Eu atoms in these compounds was indicated both from the Curie constant in the Curie-Weiss behaviors of $\chi(T)$ for $T\gg T_{\rm N}$ and also by the high-$T$ limit of the magnetic entropy $S_{\rm mag}(T)$. The $\chi(T)$ and $M(H)$ data reveal anisotropic magnetic properties for both \cusbB\ and ${\rm EuCu_2As_2}$ at $T < T_{\rm N}$.  $^{151}$Eu M\"ossbauer measurements for \cusb\ at 2.09~K confirmed that the Eu spins have a low-$T$ ordered moment of $7~\mu_{\rm B}$, consistent with the value $\mu_{\rm sat} = gS\mu_{\rm B}$ expected from $S=7/2$ and $g=2$.\cite{Ryan2015}

The anisotropic magnetic susceptibilities $\chi_{ab}$ and $\chi_c$ for ${\rm EuCu_2Sb_{1.82}}$ at $T\leq T_{\rm N}$ were modeled using molecular field theory.  Two AFM structures are equally consistent with the data.  The first is a collinear $A$-type AFM structure where the Eu ordered moments in a given $ab$~plane are ferromagnetically aligned, and the moments in neighboring layers are antiferromagnetically aligned.  In the second model the Eu ordered moments form a noncollinear planar helix with the helix axis being the $c$~axis.  In both structures the ordered moments are aligned within the $ab$~plane.  Recent neutron diffraction measurements of the magnetic structure of ${\rm EuCu_2Sb_2}$ showed that the magnetic structure is the collinear A-type structure with the ordered moments aligned in the $ab$~plane.\cite{Ryan2015}  In Sec.~\ref{Sec:MagDipoles}, it was shown that the magnetic dipole interaction favors the Eu ordered moments in the A-type AFM state to lie in the $ab$~plane, consistent with the neutron diffraction results and indicating that the magnetic dipole interaction is responsible for, or at least contributes to, determining the easy axis of the ordered moments.  On the other hand, whereas ${\rm EuFe_2As_2}$ shows the same A-type AFM structure\cite{Xiao2009} of the Eu spins as in \cusbB, Co-doped,\cite{Jin2013} P-doped,\cite{Nandi2014} and Ir-doped (Refs.~\onlinecite{Jin2015}, \onlinecite{Anand2015}) ${\rm EuFe_2As_2}$ exhibit FM order of the Eu moments with the moments oriented along the $c$~axis.  From Table~\ref{tab:MagDipole0}, FM ordering is not energetically favored compared to AFM ordering by magnetic dipole ordering and the ordered moment direction along the $c$~axis is not consistent with the easy axis predicted by this interaction.  Hence both the FM structure and the easy $c$~axis in these doped ${\rm EuFe_2As_2}$ compounds are evidently due to RKKY interactions between the Eu local moments.\cite{Jin2015} The compound ${\rm EuZn_2Sb_2}$ was reported to exhibit an AFM transition of the Eu spins at $T_{\rm N} = 13$~K.\cite{May2012}

The magnetic properties of ${\rm EuCu_2As_2}$ at $T < T_{\rm N}$ cannot be understood within molecular field theory for collinear and planar noncollinear AFM structures and instead suggest both a noncollinear and noncoplanar AFM ground state.  This type of AFM state is indicated by the strong negative curvature in $M(H)$ isotherms starting at small~$H$ for both $H\parallel c$ and $H\perp c$ in Figs.~\ref{fig:MH_EuCu2As2} and~\ref{fig:MH_EuCu2As2_2K}.

The compound ${\rm EuPd_2Sb_2}$ crystallizes in the ${\rm CaBe_2Ge_2}$-type structure, exhibits an AFM transition below 6.0~K, a spin reorientation transition near 4.5~K, and $M(H)$ isotherms with $H\perp c$ at $T = 1.8$~K, but no such field-induced transitions were observed for $H\parallel c$.\cite{Das2010}  With what we have learned about the predictions of MFT for the anisotropic $\chi(T)$ of planar noncollinear Heisenberg AFMs\cite{Johnston2011, Johnston2012, Johnston2015} since Ref.~\onlinecite{Das2010} was published in 2010, the $\chi(T)$ data in Fig.~3(a) of Ref.~\onlinecite{Das2010} for ${\rm EuPd_2Sb_2}$ strongly suggest a noncollinear AFM ground state such as a planar helical or cycloidal AFM structure.  Furthermore, from neutron diffraction measurements ${\rm EuCo_2P_2}$ with the ${\rm ThCr_2Si_2}$-type structure exhibits a (planar) helical structure with the Eu ordered moments aligned in the $ab$~plane with the helix axis being the $c$~axis, where the Co atoms were deduced to be nonmagnetic.\cite{Reehuis1992}  A comparative study of the reason for the divergent AFM structures of ${\rm EuCu_2As_2}$, ${\rm EuCu_2Sb_2}$, ${\rm EuPd_2Sb_2}$ and ${\rm EuCo_2P_2}$ using electronic structure calculations\cite{Jin2015} would be enlightening.


\acknowledgments

We thank Alan I. Goldman, Andreas Kreyssig and Dominic Ryan for helpful discussions.  This research was supported by the U.S.~Department of Energy, Office of Basic Energy Sciences, Division of Materials Sciences and Engineering.  Ames Laboratory is operated for the U.S.~Department of Energy by Iowa State University under Contract No.~DE-AC02-07CH11358.


\end{document}